\documentclass[11pt, a4paper]{article}

\usepackage[utf8]{inputenc}
\usepackage[T1]{fontenc}
\usepackage{dsfont}
\usepackage{libertine}
\usepackage{setspace}
\usepackage{soul}

\usepackage{amsthm}
\usepackage{amsmath}
\usepackage{amssymb}
\usepackage{prettyref}
\usepackage{cite}
\usepackage{braket}
\usepackage{enumitem}
\usepackage{xcolor}
\usepackage[margin=2.4cm]{geometry}
\usepackage[font={it,small}]{caption}
\usepackage{subcaption}
\usepackage{graphicx}
\usepackage[linktocpage=true]{hyperref}
\definecolor{redd}{rgb}{0.88671875,0.140625,0.328125}
\definecolor{greenn}{HTML}{0E957F}
\definecolor{titleblue}{HTML}{2B2179}
\hypersetup{
 colorlinks,
 citecolor=redd,
 linkcolor=titleblue,
 urlcolor=titleblue}
 
\newcommand{\uminus}{\raisebox{1.pt}{\scalebox{0.6}{$-$}}}
\newtheorem*{thm*}{Theorem}

\DeclareMathOperator{\Real}{Re}

\begin{document}
\begin{spacing}{1.2}

\vskip5mm

\begin{center}

{\Huge\textsc\bf Quantum Chaos in Liouville CFT}
\vskip10mm
Julian Sonner$^{1}$ and Benjamin Strittmatter$^{1,2}$\\
\vskip2.5em
{\small{\it 1) Department of Theoretical Physics, University of Geneva, 24 quai Ernest-Ansermet, 1211 Gen\`eve 4, Suisse}\\
\vskip3mm
{\it 2) School of Mathematics and Maxwell Institute of Mathematical Sciences,\\ University of Edinburgh, Edinburgh EH9 3FD, United Kingdom}}
\vskip2.5em
\tt{julian.sonner@unige.ch, b.strittmatter@ed.ac.uk}

\end{center}

\vskip10mm

\begin{abstract}

Fast scrambling is a distinctive feature of quantum gravity, which by means of holography is closely tied to the behaviour of large$-c$ conformal field theories. We study this phenomenon in the context of semiclassical Liouville theory, providing both insights into the mechanism of scrambling in CFTs and into the structure of Liouville theory, finding that it exhibits a maximal Lyapunov exponent despite not featuring the identity in its spectrum. However, as we show, the states contributing to the relevant correlation function can be thought of as dressed scramblons. At a technical level we we first use the path integral picture in order to derive the Euclidean four-point function in an explicit compact form. Next, we demonstrate its equivalence to a conformal block expansion, revealing an explicit but non-local map between path integral saddles and conformal blocks. By analytically continuing both expressions to Lorentzian times, we obtain two equivalent formulations of the OTOC, which we use to study the onset of chaos in Liouville theory. We take advantage of the compact form in order to extract a Lyapunov exponent and a scrambling time. From the conformal block expansion formulation of the OTOC we learn that scrambling shifts the dominance of conformal blocks from heavy primaries at early times to the lightest primary at late times. Finally, we discuss our results in the context of holography.
\end{abstract}

\thispagestyle{empty}
\setcounter{tocdepth}{2}

\newpage
\clearpage
\pagenumbering{arabic}

\tableofcontents

\section{Introduction}
\label{sec:intro}
The chaotic nature of a physical system is a very powerful determinant of its generic behaviour, for example whether it exhibits thermalisation or whether it retains some memory of its initial conditions, or whether it tends to evolve towards states that are close approximants to true (Haar) random behavior. These questions have found their way into the study of conformal field theories, and in particular holographic conformal field theories, largely because the holographic correspondence between gravity in AdS and certain conformal field theories exhibits the hallmarks of quantum chaos, often in a particularly strong form \cite{kitaev}. On the one hand, CFTs dual to gravity in AdS have quantum Lyapunov exponents, $\lambda=2\pi T$, which saturate the MSS bound \cite{Maldacena:2015waa}, and are thus colloquially speaking the most efficient information scramblers allowed by unitary quantum mechanics. On the other hand, two and, more tentatively three-dimensional, theories of pure gravity\footnote{That is theories which at low energies only contain gravitational degrees of freedom and no others, but may be completed at very high energy by a set of so-called black-hole microstates.} seem to satisfy some maximality principle regarding spectral notions of chaos. By this we mean that, by random-matrix universality, any chaotic theory reduces to a Wigner-type matrix integral if probed at late enough times \cite{Altland:2020ccq}, but there is now good evidence that two and three-dimensional gravity in fact reduce to such ensembles on \emph{all} time scales \cite{Saad:2019lba, Belin:2023efa}. In all of the contexts above, Liouville CFT appears to play a special role, as developed in \cite{Chandra:2022bqq,Collier:2023fwi,Belin:2023efa,Collier:2023cyw,Collier:2024mgv}. In some sense, Liouville theory seems to represent the universal chaotic kernel of all holographic 2D CFTs, and as such also holds the key to understanding the black-hole Hilbert space in AdS$_3/$CFT$_2$ duality. Of course this is not a new idea (see for example \cite{Jackson:2014nla,Turiaci:2016cvo}), and has found its manifestation in a number of works already cited, as well as underlying the Cardy-like techniques for extracting the asymptotic averaged behavior of OPE coefficients in holographic CFTs \cite{Collier:2019weq,Collier:2018exn,Belin:2021ryy,Anous:2021caj}. It is thus of obvious physical interest to further elucidate the chaotic nature of Liouville CFT itself.

In this paper we add another data point to the universal chaotic properties of Liouville CFT, by computing directly its scrambling (Lyapunov) exponent using the Liouville path integral and confirming that it saturates the MSS bound \cite{Maldacena:2015waa}
\begin{equation}
    \lambda_{\rm L}= 2\pi T\,,
\end{equation}
where here L stands for Liouville (and/or Lyapunov, in this case). We obtain this result by directly evaluating the Liouville path integral for semi-classical, that is large-$c$, heavy-heavy-light-light (HHLL) correlation functions in the thermal state, that is we compute
\begin{equation}
   G_{\rm Liouville}(z, \bar z)= \left\langle V_{\alpha_H}(\infty)V_{\alpha_H}(1) V_{\alpha_L}(z,\bar z) V_{\alpha_L}(0) \right\rangle_\beta
\end{equation}
for four Liouville vertex operators of weights $h_H, \bar h_H \sim {\cal O}(c)$ and $h_L, \bar h_L \sim {\cal O}(\epsilon c)$ for $\epsilon \ll 1$, which we then continue to a Lorentzian correlator with operators inserted out of time order (OTOC). As we explain further in the bulk of the paper, the heavy operators we consider will be close to, but below the semi-classical black-hole threshold $h_H, \bar h_H \lesssim \frac{c-1}{24}$, while the light operators will be of ${\cal O}(c)$, but hierarchically smaller than $h_H$, as above. A forteriori this place them also below the semi-classical black-hole threshold, but still far above the would-be vacuum state (which does not appear in Liovuille theory). In the language of \cite{10.1143/PTP.102.319} these are local operators which create non-normalizable states.\\

With these choices of operator dimensions, we are able to evaluate the full path integral via a saddle-point approximation, and by integrating and summing over the moduli space of contributing saddle-point solutions. The resulting expression provides a closed form for $G_{\rm Liouville}(z,\bar z)$. Using a certain hypergeometric function identity we are able to re-expand the correlation function into a sum of semi-classical conformal blocks with expansion coefficients that reproduce the semi-classical limit of the DOZZ formula, and which allows us to conclude that the exchanged operators lie on a discrete series with dimension $\Delta = 4h_L + 2r$, ($r\in \mathbb{N}_0$), which we identify with light double traces of the form $V_{\alpha_L}\partial^r V_{\alpha_L}$. It is striking that we find a maximal Lyapunov exponent in a theory which does not contain the identity, which is usually taken to be responsible for the maximal scrambling behavior. Here this behavior instead arises from a resummation of an infinite tower of double-trace exchanges. Furthermore, our closed-form analytic expressions allow us to understand the precise dominance, and exchange of dominance, of the individual blocks contributing to $G_{\rm Liouville}(z,\bar z)$, both in the Euclidean region and on the second Riemann sheet relevant for the OTOC Lorentzian ordering of operators.

We shall now give a brief summary of our main results and outline the technical steps needed to obtain them.
\subsection{Scrambling dynamics in Liouville CFT}
One of the defining properties of any physical theory is the rate at which quantum information is able to propagate, which is closely tied to the notion of quantum chaos. More precisely, chaos is realised through the process of scrambling, which can be defined as the thermalisation of an initial perturbation through the chaotic dynamics of the system. Attached to it comes a characteristic timescale known as the scrambling time. In more practical terms, this behaviour can be extracted from the growth of commutators of generic operators \cite{larkin1969quasiclassical,Maldacena:2015waa,kitaev}, which in a chaotic theory are expected to grow exponentially. In a thermal many-body system, we can define the observable

\begin{equation}
\begin{split}
	-\langle\,[V,W(t)]^2\rangle_{\beta}\;&\sim\,\,\;~~ \langle W(t)VVW(t) \rangle_{\beta}\,+\, \langle VW(t)W(t)V \rangle_{\beta}\\
	 &~~~~~\;-\langle VW(t)VW(t)\rangle_{\beta}\, -\, \langle W(t)VW(t)V \rangle_{\beta}\;\;\;,
\end{split}
\end{equation}\\[-4ex]

where $\langle\,\cdot\,\rangle_{\beta} = \text{Tr}\,[\,\rho_{\beta}\,\cdot\,]$, $W(t)$ is the Heisenberg operator $W(t)\!=\!e^{-iHt}\,W(0)\,e^{iHt}$ and $V\!:=\! V(0)$. The two terms with positive sign can be seen as expectation values of $V$ in the state $W(t)$ and vice versa, and they approach $\langle VV\rangle_{\beta}\langle WW\rangle_{\beta}$ at sufficiently late times\footnote{This is the case once $t$ grows larger than the dissipation timescale but less than the scrambling time.} regardless of the initial choice of operators $V$ and $W$. The terms with negative sign are out-of-time ordered correlators (OTOCs), and we are going to define\\[-2ex]
\begin{equation}
\label{eq:OTOCdef}
	\text{OTOC}\,:=\,\langle W(t)VW(t)V \rangle_{\beta}\;\;.
\end{equation}\\[-4ex]

In a chaotic theory, this term (along with the second out-of-time ordered piece) is expected to decay exponentially \cite{kitaev}:
\begin{equation}
	\text{OTOC} \;\sim\; \,C_0 \,-\,C_1\,e^{\lambda_L(t-t_*)}\; +\; \dots
\end{equation}\\[-4ex]

Here, $C_0$ and $C_1$ are constants. The out-of-time ordered piece captures the essence of chaotic dynamics, and is characterised by the two variables governing the exponential decay, i.e. the scrambling time $t_*$ and the Lyapunov exponent $\lambda_L$. In \cite{Maldacena:2015waa} it was demonstrated that by imposing basic requirements on the theory, such as unitarity and the late-time factorisation of the time-ordered terms, it can be shown that the Lyapunov exponent is bounded by

\begin{equation}
\label{eq:chaosb}
\lambda_L \leq \frac{2\pi}{\beta}
\end{equation}\\[-4ex]
\

Going back in the timeline, the phenomenon of scrambling has attracted considerable attention in the context of quantum gravity, when \cite{Sekino:2008he} conjectured that black holes are the most efficient scramblers in nature. Moreover, the authors anticipated that the timescale of scrambling should be logarithmic in the system size. Putting to use the AdS/CFT correspondence\cite{Maldacena:1997re,Witten:1998qj,Gubser:1998bc}, this conjecture sparked an investigation on both sides of the holographic duality. By taking advantage of the duality between the eternal black hole and the thermofield double\cite{Maldacena:2001kr}, it was shown that OTOCs computed in a BTZ black hole background are realised in terms of shock waves that distort the original geometry\cite{Shenker:2013pqa}, disrupting entanglement between the left and right boundaries. The connection to the CFT description was made in \cite{Roberts:2014ifa} (and in \cite{Caputa:2015waa}), where it was shown that the Virasoro vacuum block decays with a Lyapunov exponent saturating the bound (\ref{eq:chaosb}), and with a scrambling time logarithmic in the central charge

\begin{equation}
t_*=\frac{\beta}{2\pi} \log \frac{c}{h_w}\;\;,
\end{equation}\\[-4ex]

where $h_w$ is the holomorphic weight of the operator $W$, thus confirming the conjecture in \cite{Sekino:2008he}. Chaos in CFTs was further analysed in \cite{Fitzpatrick:2016thx,Perlmutter:2016pkf,Turiaci:2016cvo}. It was soon realised however that the OTOC in a generic CFT is not well approximated by the identity block in general, and the contribution of non-vacuum primaries was investigated in \cite{Chang:2018nzm,Liu:2018iki,Hampapura:2018otw}, in an attempt to pin down $t_*$. In this paper, we add to the discussion by presenting the example of semi-classical Liouville CFT in the heavy-light limit, which not only provides insights into the mechanism of scrambling in a CFT, but also reveals some structural properties of Liouville theory in the process. After giving a brief introduction into the path integral formulation of semiclassical Liouville CFT in Section \ref{sec:LCFT}, we show it can used to derive a compact expression for the Euclidean 4-point function in Chapter \ref{sec:HL4}. By applying a hypergeometric identity, we show that this compact result is in fact a resummed version of the conformal block expansion in Section \ref{sec:cbe}, revealing a map between path integral saddle points and conformal blocks. Moreover, the resummation provides us with the operator dimensions exchanged in semiclassical Liouville theory as well as the OPE coefficients. By means of analytic continuation we go on to derive the corresponding OTOCs for the two equivalent formulations of the heavy-light 4-point function in Chapter \ref{sec:otoc}. Equipped with two versions of the OTOC, one in compact form and one as an expansion over conformal blocks, we investigate the onset of chaos in Liouville theory in Chapter \ref{sec:chaos}. From the compact expression we are able to extract a maximal Lyapunov exponent of $\lambda_L=\frac{2\pi}{\beta}$, as well as a scrambling time $t_*=\frac{\beta}{2\pi} \log \frac{c}{h_w}$, without having to resort to any assumptions on which primary operators dominate the exchange. The OTOC in conformal block expansion form unveils that at scrambling time the hierarchy of conformal blocks gets inverted, with the sum being dominated by the heaviest operators in the spectrum at early times, but the dominance shifting to the lightest primary operator in the spectrum at late times. Finally, we discuss our results in Chapter \ref{sec:disc}, and comment on their implications on the role of Liouville theory in holography.\bigskip

\subsection{Semiclassical Liouville CFT}
\label{sec:LCFT}

The goal of this section is provide a brief introduction to semiclassical Liouville theory, and to explain how heavy-light correlation functions can be obtained from the Euclidean path integral, by means of inserting suitable vertex operators. We closely follow the reference \cite{2011JHEP...12..071H}, and refer the reader to \cite{Nakayama:2004vk} and \cite{10.1143/PTP.102.319} for more in-depth reviews of Liouville.\bigskip

The Liouville action, defined on a two-dimensional Euclidean manifold, reads
\begin{equation}
\label{eq:LVaction}
	S_{LV}\;=\;\frac{1}{4\pi}\int d^2\xi\;\sqrt{g} \left\{\partial_a\phi\,\partial_b\phi\,g^{ab} + QR\phi+4\pi\mu e^{2b\phi}\right\}\;\;,
\end{equation}

where $g_{ab}$ is known as the \textit{reference metric} and comes with a curvature scalar $R$. It is related to the so-called \textit{physical metric} via $\hat{g}_{ab}=e^{\frac{2}{Q}\phi}g_{ab}$. Moreover, $\phi$ is known as the Liouville field, and $b$ is called the Liouville coupling constant. We take the parameter $\mu$ to be positive, and $Q$ relates to $b$ as $Q=b+b^{-1}$. The resulting theory is conformally invariant, with a central charge of 
\begin{equation}
    c=1+6Q^2\;.
\end{equation}

In this language, the semiclassical limit is obtained by taking $b\rightarrow 0$, which amounts to sending the central charge to infinity. Primary fields in the theory correspond to vertex operators of the form

\begin{equation}
    V_{\alpha} = e^{2\alpha\phi}\;\;,
\end{equation}

where $\alpha$ is the Liouville momentum, and they transform with a conformal weight of

\begin{equation}
h_{\alpha} = \bar{h}_{\alpha} =  \alpha\,(Q-\alpha)\;\;.
\end{equation}\\[-5ex]

A particularly convenient description can be obtained by placing the theory on the two-sphere, choosing the flat reference metric $ds^2 =dz\,d\bar{z} $ and thus eliminating the curvature term in the action. With this choice made however, a bit of care is required in order to ensure the action remains finite. As is explained in detail in \cite{2011JHEP...12..071H} and \cite{Zamolodchikov:1995aa}, this requires the introduction of boundary terms, and additionally we need to impose boundary conditions on the field $\phi$ at infinity to guarantee a smooth physical metric. Rescaling the field as $\phi_c=2b\phi$ greatly facilitates the treatment of the semiclassical limit, and we finally obtain the semiclassical action

\begin{equation}
\label{eq:SLV}
	b^2\;S_{LV}\;=\;\frac{1}{16\pi}\int_{D} d^2\xi\;\left\{\left(\partial_a\phi_c\right)^2+16\pi\mu b^2 e^{\phi_c}\right\} \;+\; \frac{1}{2\pi}\oint_{\partial D}\;\phi_c d\theta + 2\log R + \mathcal{O}(b^2)\;\;.
\end{equation}\\[-5ex]

The domain of integration is regulated by choosing to integrate over a disk $D$ of radius $R$, which is later taken to be infinite. Finally we require that the field obeys the boundary condition at infinity

\begin{equation}
    \phi_c(z,\bar{z}) = -2\log(z\bar{z}) + \mathcal{O}(1)\;\;\;\;\;\text{as}\;\;\;\;\;|z|\rightarrow \infty \;\;.
\end{equation}\\[-5ex]

Having taken these precautions, with (\ref{eq:SLV}) we have achieved a well defined formulation of semiclassical Liouville theory on the two-sphere. The equation of motion then reads

\begin{equation}
\label{eq:eom}
	\partial\bar{\partial}\phi_c = 2 \lambda e^{\phi_c}\;\;,
\end{equation}\\[-5ex]

where we have defined the parameter $\lambda = \pi\mu b^2$, which is held fixed in the semiclassical limit. Solutions to this equation correspond to saddles of the path integral, which can be straightforwardly constructed from the action (\ref{eq:SLV}). The idea now is to build correlation functions by inserting operators into the path integral, and to evaluate the resulting integral as a sum over its saddle points. In order to do this correctly, it is important to distinguish between two types of operators: Heavy operators $V_{\alpha_H}$ are characterised by a Liouville momentum scaling as $\alpha_{H} = \eta_H/b$, and their presence causes a backreaction on the geometry that deforms the saddle points. In \cite{2011JHEP...12..071H} it is explained how the effect of heavy operators can be incorporated into a modified action $\tilde{S}_{LV}$, which we briefly review in Appendix \ref{sec:inttrick}, but we will not need its explicit form here. Light operators $V_{\alpha_L}$ in contrast have no influence on the shape of saddles, as their momenta scale as $\alpha_L = \sigma_L \,b$, where we take the parameters $\eta_H$ and $\sigma_L$ to be fixed in the semiclassical limit. In this setup, the light operators can be interpreted as probes that propagate on a background shaped by the heavy insertions. Let us denote these two operator types by

\begin{equation}
\label{eq:weights2}
   \tilde{H}(z,\bar{z}) \equiv V_{\eta_H/b}\,(z,\bar{z})
~~~~~~~~~~~\text{and}~~~~~~~~~~~
    \tilde{L}(z,\bar{z}) \equiv V_{\tilde{\sigma}_L b}\,(z,\bar{z})
\end{equation}\\[-5ex]

with conformal weights scaling as $ \tilde{h}_H \sim c$ and  $ \tilde{h}_L\sim \mathcal{O}(1)$. Now, we have all the building blocks needed, and we can assemble the semiclassical correlation function of $m$ heavy and $n$ light operators as

\begin{equation}
\label{eq:semicor}
	\langle V_{\alpha_{H,1}}(z_1,\bar{z}_1)\dots V_{\alpha_{H,m}}(z_m,\bar{z}_m)V_{\alpha_{L,1}}(x_1,\bar{x}_1)\dots V_{\alpha_{L,n}}(x_n,\bar{x}_n )\rangle \approx \sum_{\rm saddles}e^{-\tilde{S}_{LV}} \; \prod_{i=1}^{n}\;e^{\alpha_{L,i}\phi_c(x_i,\bar{x}_i)/b}\;\;.
\end{equation}\\[-3ex]
Here it is crucial that the right hand side sums over all possible saddle points, if there are multiple solutions $\phi_c$ to the equation of motion (\ref{eq:eom}). We shall make this explicit in the next section, where we are going use the semiclassical formula to compute the heavy-light four point function.\bigskip

In this paper, we work with a slightly different hierarchy of operator dimensions, as in \cite{Fitzpatrick:2014vua,Anous:2016kss}, whereby we tweak the above scaling dimensions of operators \cite{2017JHEP...08..045B}, in order to obtain a more interesting limit from the gravity perspective. Instead of keeping $h_L$ constant in the semiclassical limit, instead we also take it to scale with the central charge. This can be achieved by relabelling $\sigma_L = \eta_L/b^2$, and demanding that $\eta_L$ remains fixed as $b\rightarrow 0$. In order to distinguish between heavy and light states, we take $\eta_L$ to be parametrically small ($\eta_L \ll 1$), whereas $\eta_H$ remains of order $\eta_H\sim\mathcal{O}(1)$.  To summarise, we have heavy and light operators and their respective conformal weights given by 

\begin{equation}
\label{eq:weights}
\begin{split}
   &H(z,\bar{z}) \equiv V_{\eta_H/b}\,(z,\bar{z})
~~~~~~~~~~~\text{with}~~~~~~~~~~~h_H=\,\frac{\eta_H}{b^2}(1-\eta_H) \,\sim\; c\\
   & 
 L(z,\bar{z}) \equiv V_{\sigma_L b}\,(z,\bar{z})~~~~~~~~~~~~\;\text{with}~\,~~~~~~~~~~
    h_L=\;\sigma_L\;=\, \frac{\eta_L}{b^2}\,\sim\; c~~~~~~~~~~~~.
\end{split}
\end{equation}\\[-5ex]

Following \cite{2017JHEP...08..045B}, we will from now on only consider the 'holographic' tweak of the heavy-light limit, and make the assumption that the formalism outlined in \cite{2011JHEP...12..071H}, and in particular equation (\ref{eq:semicor}), still give an accurate description in this regime. As we show in Section \ref{sec:cbe}, the conformal block expansion of semi-classical correlations between heavy and light operators exchanges a discrete tower of spinless, double-trace operators of weight

\begin{equation}
    h=\bar{h}=2h_L+r~~, ~~~ r\in \mathbb{N}_0\;.
\end{equation}

Another feature we shall observe is that in order for the correlations to be non-vanishing, the conformal weights of the external heavy operators are forced to lie in the discrete set

\begin{equation}
\eta_H=\frac{1}{2}+\frac{kb^2}{2}~~,~~k\in\mathbb{Z}~~.
\end{equation}\\[-5ex]

We shall later comment on the Seiberg bound \cite{10.1143/PTP.102.319}, which selects negative values of k. In the next section, we are going to see how the heavy-light four point function can be explicitly evaluated as a sum over the saddles of the semi-classical action.\bigskip

\section{Four-Point Function from Path Integral}
\label{sec:HL4}
Having outlined how semiclassical heavy-light correlation functions can be obtained from the path integral, we are now ready to compute the four-point function. We are interested in the special case of two identical heavy operators and two identical light operators. Given the prescription (\ref{eq:semicor}), it is straightforward to write down the corresponding expression:

 \begin{equation}
 \label{eq:4pt}
 	\left\langle H_{\frac{\eta_H}{b}}(z_1,\bar{z}_1)\,  H_{\frac{\eta_H}{b}}(z_2,\bar{z}_2)\, L_{\sigma_L b}(z_3,\bar{z}_3)\, L_{\sigma_L b}(z_4,\bar{z}_4)  \right\rangle \;\approx \; \sum_{\text{saddles}} e^{-\tilde{S}_{LV}[\phi_{\eta_{H}}]} \; e^{\sigma_L \,\phi_{\eta_H}(z_3,\bar{z}_3)} \; e^{\sigma_L\, \phi_{\eta_{H}}(z_4,\bar{z}_4)}\;.
 \end{equation}\\[-4ex]
 
 The exponential factor involving $\tilde{S}_{LV}$ represents the heavy insertions, and their effect on the correlation function is packaged into a renormalised modified action, evaluated on a given solution $\phi_{\eta_H}$ of the equations of motion. The two light insertions on the other hand produce the conventional factors corresponding to the standard vertex operators of the theory. The subscript of the field $\phi_{\eta_H}$ reflects the fact that the equations of motion are only sensitive to the presence of the heavy operators. In order to determine which saddles we should sum over, let us first state the result for both the fields and the modified action. In the presence of two identical heavy operators, a solution to the equations of motion was found in \cite{2011JHEP...12..071H}:
 
  \begin{equation}
 \label{eq:field}
 	e^{\phi_{\eta_H}} \; = \; \frac{1}{\lambda} \frac{\kappa^2}{\;\left(\;\kappa^2\, |z-z_1|^{2\eta_H}\, |z-z_2|^{2-2\eta_H} - \frac{1}{(1-2\eta_H)^2\,|z_{12}|^2}\,|z-z_1|^{2-2\eta_H}\,|z-z_2|^{2\eta_H}\;\right)^2}\;\;
 \end{equation}\\[-4ex]
 
 Up to a power of  $\sigma_L$, this corresponds to the vertex operators of the two light operator insertions. The parameter $\kappa$ is an arbitrary complex number such that the denominator on the right hand side is non-vanishing. Because the expression depends only on $\kappa^2$, the moduli space associated with this solution is isomorphic to the complex half plane $\mathbb{H}$. The prescription (\ref{eq:4pt}) for the correlation function therefore instructs us to integrate over $\kappa$. Next, let us take a closer look at the contribution coming from the heavy operators. The modified action can be evaluated using an integration trick (proposed in \cite{Zamolodchikov:1995aa} and summarised in Appendix \ref{sec:inttrick}), and reads
 
 \begin{equation}
 \label{eq:modS}
 \begin{split}
 	b^2\tilde{S}_{LV}\; &= \; 2 \pi i \left(N + 1/2\right) (1 - 2\eta_H) + (2\eta_H -1)\log\lambda + 4(\eta_H-\eta_H^2) \log |z_{12}|\\
 		&\;\;+2\left[(1-2\eta_H) \log (1-2\eta_H)-(1-2\eta_H) \right]\;\;.
 \end{split}
 \end{equation}\\[-4ex]
 
 The parameter $N$ is an integer, which labels the choice of branch when taking the logarithm of (\ref{eq:field}), and represents another degree of freedom in the solution that needs to be summed over. Note that $\eta_H<1/2$ by assumption (see appendix), so the logarithm is single valued. To summarise,  the resulting moduli space of saddles, defined by $\kappa$ and $N$, becomes $\mathbb{H}\times \mathbb{Z}$. With this information, we can start to assemble the RHS of (\ref{eq:4pt}).
 
 \begin{equation}
 \label{eq:sumsaddles}
 	\left\langle H(z_1,\bar{z}_1)\,  H(z_2,\bar{z}_2)\, L(z_3,\bar{z}_3)\, L(z_4,\bar{z}_4) \right\rangle\; = \; \sum_{N \in \mathbb{Z}} e^{-\tilde{S}_{LV}[\phi_{\eta_{H}}]}\;\int_{\mathbb{H}} d\kappa  \wedge d \bar{\kappa} \;\frac{\kappa^{4\sigma_L}}{(\zeta_3\kappa^2- \chi_3 \;)^{2\sigma_L}(\zeta_4\kappa^2- \chi_4\;)^{2\sigma_L}}
 	,
 \end{equation}\\[-4ex]

 where we have introduced the following shorthand notation\footnote{This notation was introduced in \cite{2017JHEP...08..045B}, however this reference does not evaluate the expression (\ref{eq:sumsaddles}), which is crucial for the application to chaos correlators in this work.}:
 
 \begin{equation}
     \zeta_i = |z_{1i}|^{2\eta_H}|z_{2i}|^{2-2\eta_H} ~~~~~~~~~~\text{and}~~~~~~~~~~~\chi_i = \frac{|z_{1i}|^{2-2\eta_H} \,|z_{2i}|^{2\eta_H}}{(1-2\eta_H)^2 |z_{12}|^2}
 \end{equation}\\[-4ex]

 Here we are making use of the standard notation $|z_{ij}|=|z_i-z_j|$. We notice that the integrand in (\ref{eq:sumsaddles}) has no dependence on $N$, so the sum over $\mathbb{Z}$ and the integral are factors that can be evaluated independently, and we denote them respectively by $\langle HHLL\rangle=\mathcal{H}\times\mathcal{Z}$.
Note that the ratio
\begin{equation}
    \frac{\chi_3 \zeta_4}{\zeta_3 \chi_4} = |1-z|^{4\eta_H-2} = |1-z|^{-2\tilde\alpha}\,,
\end{equation}
where $z$ is the conformal cross-ratio, which will become useful below:
\begin{equation}
    z = \frac{z_{12}z_{34} }{z_{13} z_{24}}
\end{equation}

\subsection{Sum Over Saddles}
\label{sec:sumsad}

In this section we show how the sum over semiclassical saddles in the expression for the heavy-light four point function (\ref{eq:sumsaddles}) can be evaluated explicitly\footnote{ closely following the treatment in the Master's Thesis of B. Strittmatter \cite{MSc}.}. First, we deal with the integral over the continuous sector of moduli space. As suggested in \cite{2017JHEP...08..045B}, the integral in (\ref{eq:sumsaddles}) can be rewritten using Cauchy's integral formula and Stokes' theorem as an integral over the boundary of $\mathbb{H}$, where the non-vanishing contribution comes from the contour along the real line.

 \begin{equation}
 	\label{eq:modint}
  	\mathcal{H}\;\equiv\;\int_{-\infty}^{\;\infty} d\kappa \; \frac{\;\kappa^{4\sigma_L +1 }}{(\zeta_3\kappa^2- \chi_3 \;{+ \;i \varepsilon })^{2\sigma_L}(\zeta_4\kappa^2- \chi_4\;{+ \;i \varepsilon})^{2\sigma_L}}\;\;.
 \end{equation}\\[-4ex]

Here we regulated the two poles by introducing an $i\varepsilon$-description. We now propose the following way of evaluating this integral: The trick is to notice that it is a special case of the integral representation of the hypergeometric function: 

 \begin{equation}
 \label{eq:hypint}
 	{}_2F_1(a,\,b\,;\,c;\,x)\;=\;\frac{\Gamma(c)}{\Gamma(b)\Gamma(c-b)}\;\int_0^1\;dt\; t^{b-1}(1-t)^{c-b-1}(1-xt)^{-a}\;\;.
 \end{equation}\\[-4ex]
 
 In Appendix \ref{sec:int} we demonstrate the equivalence of integrals (\ref{eq:modint}) and (\ref{eq:hypint}) up to prefactors explicitly, where we identify $a=b= 2\sigma_L$, $c=4\sigma_L$ and $x = 1-\chi_3 \zeta_4/ \chi_4\zeta_3$. The result including all prefactors can then be summarised as

 \begin{equation}
 \label{eq:intres}
     \mathcal{H}= \;\frac{\omega(\sigma_L)\;\;}{(\zeta_3\;\chi_4)^{2\sigma_L}}\;\,{}_2F_1\Big(2\sigma_L,\, 2\sigma_L;\, 4\sigma_L;\, 1-\frac{\chi_3 \zeta_4}{ \chi_4\zeta_3}\Big)\;,
 \end{equation}\\[-4ex]

 where we have grouped together all coordinate-independent coefficients into the constant
 
 \begin{equation}
     \omega(\sigma_L)\;=\;\frac{1}{2}\left(e^{\pi i+ 2\pi i K}\right)^{2\sigma_L}\left(1+\left(e^{\pi i+2\pi i M}\right)^{4\sigma_L+1}\right)\frac{\Gamma(2\sigma_L)^2}{\Gamma(4\sigma_L)}\;.
 \end{equation}\\[-4ex]
 
 \bigskip

 Next, we proceed by evaluating the discrete part of moduli space $\mathcal{Z}$ in (\ref{eq:sumsaddles}). Plugging the modified action (\ref{eq:modS}) into the expression, and defining the shorthand \footnote{Note that together with (\ref{eq:weights}), this implies that $\tilde{\alpha}=\sqrt{1-\frac{24 h_H}{c}}$ in the semiclassical limit.} $\tilde{\alpha}\equiv 1-2\eta_H$, we find that the sum evaluates to

 \begin{equation}
 \label{eq:sumres}
 \mathcal{Z} =\;\lambda^{\tilde{\alpha}/b^2}\;\tilde{\alpha}^{{-2\tilde{\alpha}/b^2}}\;\;e^{2\tilde{\alpha}/b^2}\;|z_{12}|^{-4h_H}\; \sum_{N\in\mathbb{Z}}\;e^{-\frac{2\pi\,i}{b^2}\,\tilde{\alpha}\,(N+\frac{1}{2})}
 \end{equation}\\[-4ex]
 
  Interesting here is the sum over oscillating phases, which in \cite{2011JHEP...12..071H} defined a Stokes line in parameter space, i.e. in the space that is defined by the complex parameters $\eta_H$ and $b$. Said Stokes wall then sits on the line $\text{Im}\,(\tilde{\alpha})=\text{Im}\,((2\eta_H-1)/b^2)=0$, and dictates which set of saddles needs to be summed over. In our case however, both parameters are real. Nevertheless, the Stokes line imposes a meaningful constraint on the theory in this case, which can be seen by applying Poisson's summation formula:
 
\begin{equation}
\label{eq:deltaid}
 \sum_{N \in \mathbb{Z}} e^{-2\pi i\,N\,\tilde{\alpha}/b^2}\;=\; \sum_{k \in \mathbb{Z}} \;\delta\!\,\left(\;\frac{\tilde{\alpha}}{\,b^2}+k\;\right)
 \end{equation}\\[-4ex]
 
 For completeness, we give a summary of Poisson's theorem in Appendix \ref{sec:poisson}. We postpone our comments on the appearance of this delta-comb until after the next section, in which we derive an alternative expression for the four-point function. 
 The final expression for the four-point function is easily obtained by plugging the results (\ref{eq:intres}) and (\ref{eq:sumres}) back into (\ref{eq:sumsaddles}), using the identity (\ref{eq:deltaid}) and restoring explicit coordinate dependence \cite{MSc}:
 
 \begin{equation}
 \label{eq:4ptcc}
     \langle HHLL\rangle = \;\tilde{\omega}(h_L,\tilde{\alpha},b)\;\, \frac{|z|^{4h_L}}{|z_{12}|^{4h_H}|z_{34}|^{4h_L}}\;|1-z|^{-2h_L(\tilde{\alpha}+1)}\; {}_2F_1\left(2h_L,\, 2h_L;\, 4h_L;\, 1-\frac{1}{\;|1-z|^{2\tilde{\alpha}}}\right)\;
 \end{equation}\\[-4ex]

 We have introduced the conformal cross ratio $z=z_{12}z_{34}/z_{13}z_{24}$ and again, we have absorbed all constants into a prefactor $\tilde{\omega}$, which is given by 
 
 \begin{equation}
     \tilde{\omega}(h_L,\tilde{\alpha},b)\; =\, \omega(\sigma_L) \;\lambda^{\tilde{\alpha}/b^2}\;\tilde{\alpha}^{{4h_L-2\tilde{\alpha}/b^2}}\;\;e^{(2-\pi\,i)\tilde{\alpha}/b^2}\;\sum_{k \in \mathbb{Z}} \;\delta\!\,\left(\;\frac{\tilde{\alpha}}{\,b^2}+k\;\right)
 \end{equation}\\[-4ex]
 
 For reasons that will become apparent in the next section, we apply a hypergeometric transformation\footnote{For instance identity (15.8.1) in \cite{822801} for the choice of argument $y=1-|1-z|^{-2\tilde{\alpha}}$.} to (\ref{eq:4ptcc}) in order to bring the expression to a more familiar form:

  \begin{equation}
  \label{eq:4ptc}
     \langle HHLL\rangle = \;\tilde{\omega}(h_L,\tilde{\alpha},b)\;\, \frac{|z|^{4h_L}}{|z_{12}|^{4h_H}|z_{34}|^{4h_L}}\;|1-z|^{2h_L(\tilde{\alpha}-1)}\; {}_2F_1(2h_L,\, 2h_L;\, 4h_L;\, 1-|1-z|^{2\tilde{\alpha}})\;
 \end{equation}\\[-4ex]
 
 To summarise, in this section we succeeded in deriving the Liouville heavy-light four-point function from a path integral picture. Not only is the correlator (\ref{eq:4ptc}) exact in the semiclassical limit, but more importantly it is an explicit compact expression, which is one of the key technical results in this paper and will play a central role in the analysis of OTOCs in chapter \ref{sec:chaos}. Note that (\ref{eq:4ptc}) has the expected coordinate dependence for a conformal four-point function, and it turns out that it is a re-summed version of the conformal block expansion. We shall make this explicit in the next section.\smallskip

\subsection{Conformal Block Expansion}
\label{sec:cbe}

In the previous section we have derived a compact expression for the Euclidean heavy-light four point function. The hypergeometric function appearing in (\ref{eq:4ptc}) is reminiscent of the semiclassical heavy-light conformal block derived in \cite{2015JHEP...11..200F}, which reads:

\begin{equation}
\label{eq:blocklit}
    \mathcal{F}(h_L,h_H,h,z)\; =\; (1-z)^{(\tilde{\alpha}-1)h_L}\,\left(\frac{1-(1-z)^{\tilde{\alpha}}}{\tilde{\alpha}}\right)^{h-2h_L}\;{}_2F_1(h,\,h;\,2h;1\!-\!(1\!-\!z)^{\tilde{\alpha}})\; 
\end{equation}\\[-4ex]

As we shall see in this section, this is not a coincidence. The compact form (\ref{eq:4ptc}) of the correlator is a conformal block expansion in disguise. In its current form however, there is one crucial difference between (\ref{eq:4ptc}) and (\ref{eq:blocklit}): The hypergeometric factor in (\ref{eq:4ptc}) includes both holomorphic and anti-holomorphic coordinate dependence whereas the full conformal block appearing in the expansion factorises into a holomorphic  (\ref{eq:blocklit}) and anti-holomorphic component, i.e. contributes an overall term of $\mathcal{F}(h_L,h_H,h,z)\mathcal{F}(\bar{h}_L,\bar{h}_H,\bar{h},\bar{z})$ for each conformal primary in the sum. In order to prove equivalence between the two expressions, we need a relation of the form

\begin{equation}
    {}_2F_1\big(1-|1-z|^{2\tilde{\alpha}}\big)\;=\;\sum_{primaries} \;\; {}_2F_1\big(1-(1-z)^{\tilde{\alpha}}\big)\times {}_2F_1\big(1-(1-\bar{z})^{\tilde{\alpha}}\big)\;\;,
\end{equation}

where we have suppressed all but the last argument of the hypergeometric function for simplicity. As it turns out, this relation follows directly from the properties of the hypergeometric function. In \cite{10.1093/qmath/os-11.1.249} it was shown that

 \begin{equation}
 \label{eq:hypid}
 \begin{split}
 	{}_2F_1(a,\,b;\,c;\,x+y-xy) \; &= \sum_{r=0}^{\infty}\;(-1)^r\;\frac{(a)_r\,(b)_r\, (c-a)_r\,(c-b)_r}{r!\,(c+r-1)_r\,(c)_{2r}}\;x^ry^r {}_2F_1(a+r,\,b+r;\,c+2r;\,x)\,\\[-1.5ex]
 	&~~~~~~~~~~~~~~~~~~~~~~~~~~~~~~~~~~~~~~~~~~~~~~~~~~~~~~~~~~~~~~~~~~~~~~\times {}_2F_1(a+r,\,b+r;\,c+2r;\,y)\;.
 \end{split}
 \end{equation}

 Here, $(n)_i$ denotes the Pochhammer symbol. In order to obtain the conformal block expansion for the correlation function (\ref{eq:4ptc}), we apply the identity for $x=1-(1-z)^{\tilde{\alpha}}$ and $y=1-(1-\bar{z})^{\tilde{\alpha}}$, and find
 
\begin{equation}
\label{eq:4ptexp}
\begin{split}
     \langle HHLL\rangle &= \;\, \frac{|z|^{4h_L}}{|z_{12}|^{4h_H}|z_{34}|^{4h_L}}\;\;\sum_{h}\;C(h,h_L,h_H,c)\;\,\left|1-z\right|^{2h_L(\tilde{\alpha}-1)}\;\left|\frac{1\!-\!(1\!-\!z)^{\tilde{\alpha}}\,}{\tilde{\alpha}}\right|^{2h-4h_L}\; \;.\\
     &~~~~~~~~~~~~~~~~~~~~~~~~~~~~~~~~~~~~\times\;{}_2F_1\big(h,\, h;\,2h;\, 1-(1-z)^{\tilde{\alpha}}\big)\;{}_2F_1\big(h,\, h;\, 2h;\, 1-(1-\bar{z})^{\tilde{\alpha}}\big)\;
\end{split}
\end{equation}\\[-4ex]

We have introduced the notation $h=2h_L+r$. Again, the coefficient $C(h,h_L,h_H,c)$ groups together all coordinate-independent constants, and corresponds to the product of the two OPE coefficients in the conformal block expansion. It reads

\begin{equation}
\label{eq:OPEs}
    C(h,h_L,h_H,c)\;=\;(-1)^{h-2h_L}\,\tilde{\omega}\,\tilde{\alpha}^{2h-4h_L}\;\frac{\Gamma(4h_L)}{\Gamma(2h)\Gamma(2h-1)}\;\frac{\Gamma(h)^4}{\Gamma(2h_L)^4}\; \frac{\Gamma(h+2h_L-1)}{\Gamma(h-2h_L+1)}\;\;.
\end{equation}\\[-4ex]

In other words, with (\ref{eq:4ptexp}) we have obtained the expected conformal block expansion for a semiclassical heavy-light four-point function \cite{MSc}

\begin{equation}
\label{eq:4ptcbe}
     \langle HHLL\rangle = \;\, \frac{1}{|z_{12}|^{4h_H}|z_{34}|^{4h_L}}\;\;\sum_{h}\;C(h,h_L,h_H,c)\;z^{2h_L}\bar{z}^{2h_L}\;\left|\,\mathcal{F}(h_L,h_H,h,z)\,\right|^{2}\; \;,\\
\end{equation}\\[-4ex]

where the Virasoro conformal blocks $\mathcal{F}(h_L,h_H,h,z)$ are given by (\ref{eq:blocklit}). Moreover, it is now straight-forward to read off the spectrum of intermediate states. We find that the semi-classical limit of Liouville only exchanges a discrete set of primary operators. We find that the exchanged operators constitute an infinite tower of spin-zero states with conformal weights

\begin{equation}
\label{eq:spec}
    h = \bar{h} = 2h_L + r~~~,~~~\text{for}~~ r \in \mathbb{N}_0~,
\end{equation}\\[-4ex]

\begin{figure}
    \centering
    \includegraphics[width=0.67\textwidth]{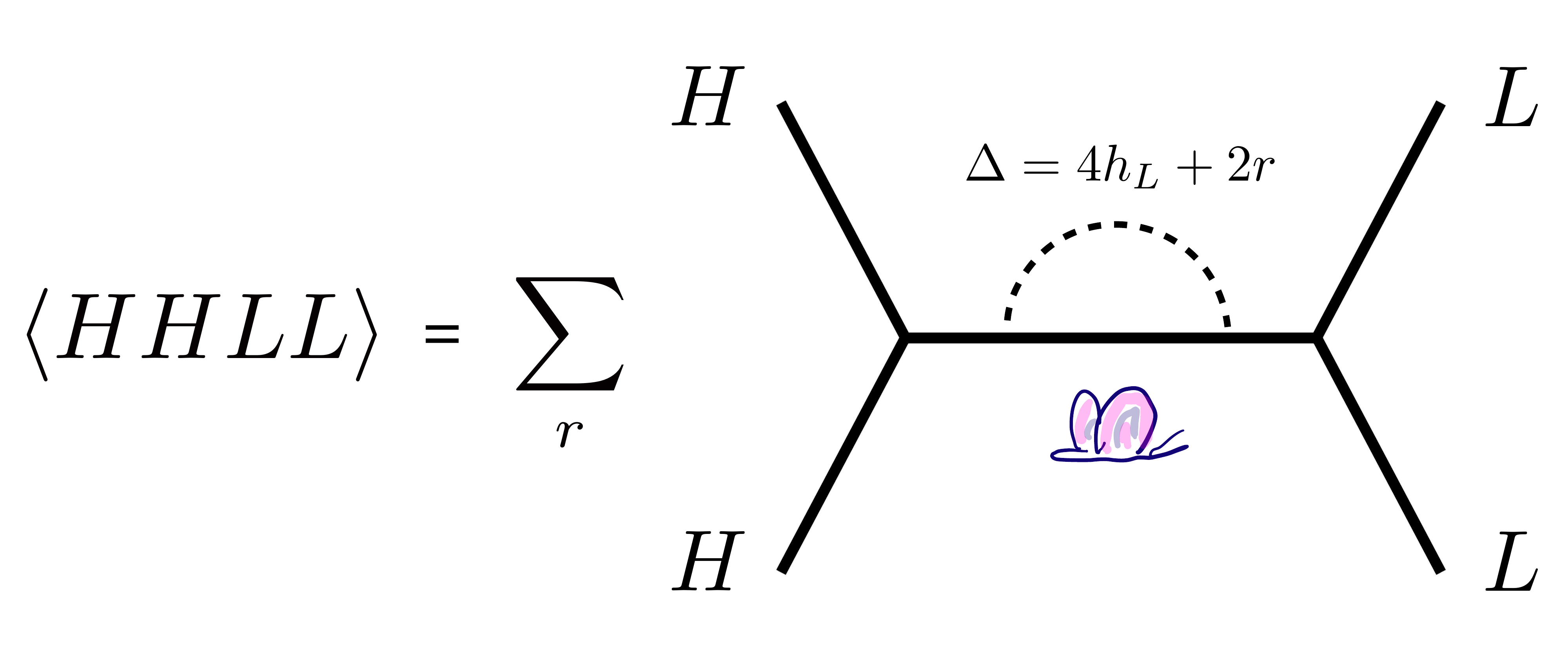}
    \caption{Schematic depiction of the conformal block expansion of the Liouville heavy-light correlator. An infinite tower of spinless double trace operators of dimension $\Delta=4h_L+2r$ is exchanged. While the identity, the so-called `scramblon' mode, is not exchanged itself, we can view each exchanged mode as a sort of dressed scramblon, since its conformal block explicitly includes the idenity as a factor (see Eq. (\ref{eq:blockfact}) and equivalently (\ref{eq:blocklit})). This effectively leads to a maximal scrambling exponent, a mechanism that contrasts with the usual identity/scramblon-exchange picture \cite{Roberts:2014ifa, kitaev}.}
    \label{fig:ZZdiag}
\end{figure}

corresponding to double-trace operators of dimension $\Delta=4h_L+2r$, of the schematic form $:
~V_{\alpha_L}\partial ^r V_{\alpha_L}~:$. Figure \ref{fig:ZZdiag} illustrates how the pair of light (heavy) operators fuses into the OPE, which in mathematical terms is embodied in the hypergeometric identity (\ref{eq:hypid}).
This is in agreement with the findings in \cite{2017JHEP...08..045B}, where the OPE was obtained by gluing together three-point functions given by the DOZZ formula. Here we give a complementary derivation of this result, which follows directly from a path integral prescription. It has long been anticipated that saddle points of the path integral correspond to the conformal blocks of the theory\cite{1969JETP...28.1200L,Anous:2017tza}, but here we demonstrate by means of a direct computation how the conformal block expansion arises in semiclassical Liouville theory.  In particular, our calculation shows that the correspondence is non-local, in the sense that only the sum over all conformal blocks corresponds to the integral over the continuous moduli space of the semiclassical saddle points, but individual saddles cannot be identified with individual conformal blocks. Another peculiar feature of our calculation is the appearance of the delta-comb (\ref{eq:deltaid}) found at the end of section \ref{sec:sumsad}. It suggests that the correlation function is non-vanishing only for a discrete set of points. Therefore, the comb imposes a "quantisation condition" on the external operators, which have to obey the following relation:

\begin{equation}
 \label{eq:etah}
 	\eta_H\;=\;\frac{1}{2}+\frac{k\,b^2}{2}\;\;\;\;\;\;\;,\;\;\;k\in \mathbb{Z}\;.
 \end{equation}\\[-4ex]

\begin{figure}
    \centering
    \includegraphics[width=0.67\textwidth]{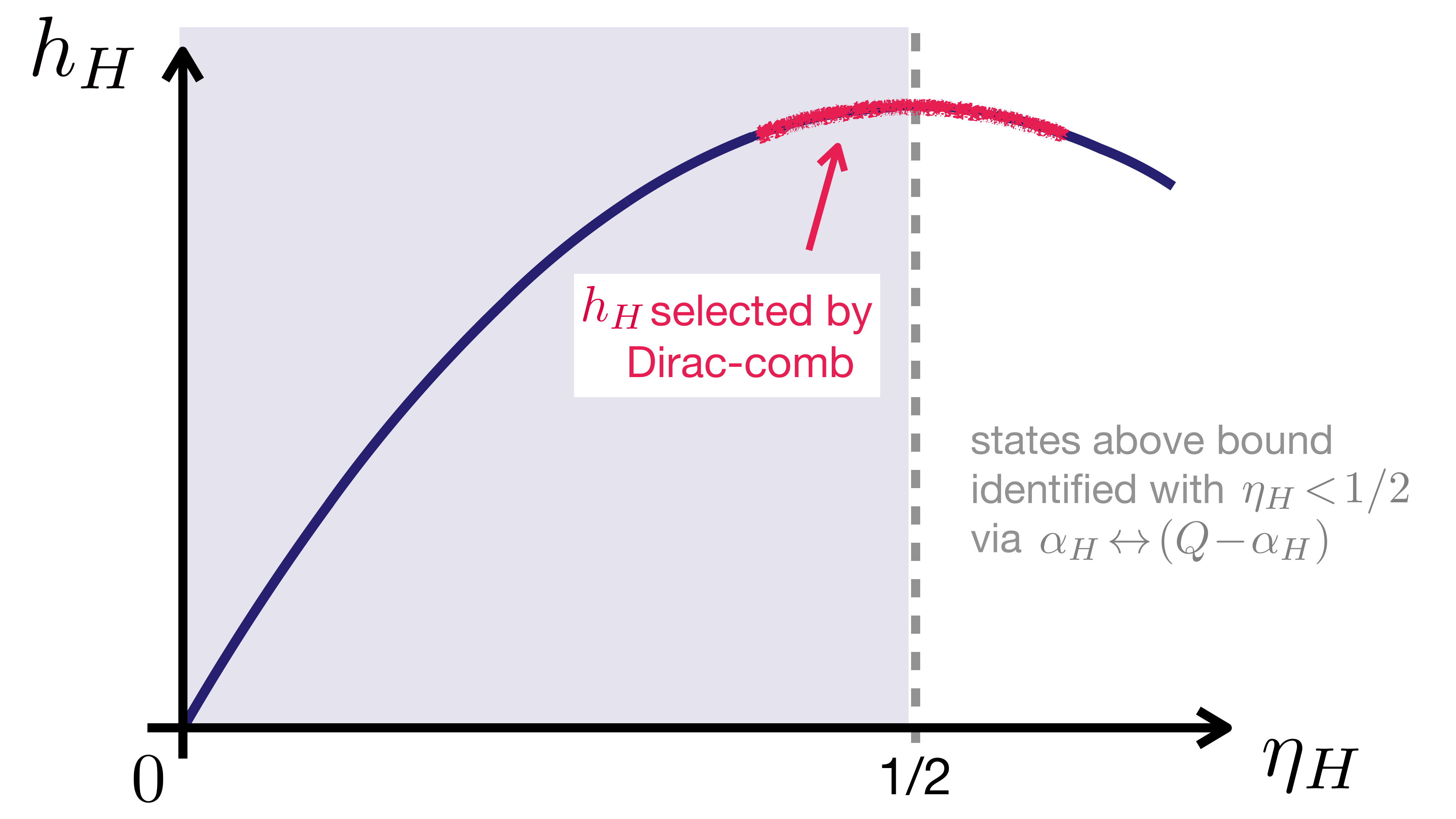}
    \caption{The conformal weights $h_H$ of the external heavy operators lie on a curve $h_H=\frac{\eta_H}{b^2}(1-\eta_H)$. The Dirac comb (\ref{eq:deltaid}) implies condition (\ref{eq:etah}), which sets $\eta_H\sim\frac{1}{2}+ \mathcal{O}(b^2)$, and hence dictates that the heavy operators are chosen close to the maximum permitted value. This places them near the black-hole threshold $h_{H}^{BH}\sim\frac{c-1}{24}$, which coincides with the Seiberg bound $\eta_H=\frac{1}{2}$.}
    \label{fig:weights}
\end{figure}

Interestingly, this places the heavy operators very close to the Seiberg\footnote{As explained in \cite{2011JHEP...12..071H}, we need not worry about violating the bound for positive $k$. Because  Liouville momenta $\alpha$ and $(Q-\alpha)$ correspond to the same quantum state, one can always pick one of two operators that satisfies the bound.} bound $Re(\eta)<\frac{1}{2}$, as we are operating in the regime $b\to 0$. The correlator instructs us to set the external heavy operator weights close to the maximum permitted by the bound, hinting that semi-classical scattering in Liouville knows about aspects of the underlying quantum theory. We further note that the Seiberg bound coincides with the black hole threshold, so (\ref{eq:etah}) ensures that the heavy operators lie close to but below threshold $h_H\lesssim\frac{c-1}{24}$. Moreover, this condition resonates with the fact that the spectrum of exchanged operators (\ref{eq:spec}) is discrete, in contrast to the a priori continuous spectrum in Liouville theory. The appearance of a delta function enforcing this may seem odd, but they are actually a common occurrence in Liouville CFT. Not only are delta functions present in the two point function of two identical operators, but also make an appearance in the DOZZ formula\cite{2011JHEP...12..071H}. The fact that only the heavy external states are subject to this constraint can be attributed to the way we set up the path integral formalism in (\ref{eq:semicor}), where we treated the light operators as probes that do not alter the geometry.

In summary, in this section we have derived two equivalent expressions for the semiclassical correlation function in the heavy-light limit, once in compact form (\ref{eq:4ptc}) and once as a conformal block expansion (\ref{eq:4ptcbe}). Equipped with these results, we would now like to inspect the chaotic properties of Liouville theory. This requires knowledge of the out-of-time ordered correlator, and in the next section we are going to show how they can be obtained from the Euclidean expressions by means of analytic continuation.\bigskip

\section{Out-Of-Time Order Correlator}
\label{sec:otoc}
In this section we give a short summary of how out-of-time ordered correlators can be obtained through analytic continuation of their Euclidean counterparts in 2D CFT, following the prescription in \cite{Roberts:2014ifa}. We then go on to calculate the OTOCs corresponding to the two results (\ref{eq:4ptc}) and (\ref{eq:4ptcbe}) in the previous section. 

A correlation function can be thought of as a multi-sheeted Riemann surface, where the principal sheet corresponds to the Euclidean configuration, while the other sheets capture the Lorentzian correlators, each with their unique choice of ordering of operators \cite{Osterwalder:1973dx}. Due to Wightman's theorem, each of these correlators is an analytic continuation of the Euclidean one, which can be exploited in order to derive the OTOC. 
Our starting point will be our analytical expression

\begin{equation}
\label{eq:form}
 	\big\langle H(z_1,\bar{z}_1)H(z_2,\bar{z}_2)L(z_3,\bar{z}_3)L(z_4,\bar{z}_4)\big\rangle_{\beta} \; =\; \frac{1}{|z_{12}|^{4h_H}|z_{34}|^{4h_L}\!\!}\;f(z,\bar{z})\;\;,
\end{equation}\\[-4ex]

for the correlation, with explicit expressions for the function $f(z,\bar z)$ given by (\ref{eq:4ptc}) and (\ref{eq:4ptcbe}). We can move from the Euclidean to the Lorentzian sheets by assigning a specific order of operators, which can be done through the following prescription. First, we map the above CFT to a cylinder, using the conformal transformation

\begin{equation}
\label{eq:zmap}
    z_i = \exp\Big[\frac{2 \pi}{\beta}(x_i + t_i)\Big]~~~~~~~~\text{and}~~~~~~~~~\bar{z}_i = \exp\Big[\frac{2 \pi}{\beta}(x_i - t_i)\Big]\;\;.
\end{equation}\\[-4ex]

The resulting system is now a thermal one at inverse temperature $\beta$. In this convention, real $t$ corresponds to Lorentzian and imaginary $t$ to Euclidean time. Next, we assign a small imaginary time $t_j= i \epsilon_j$ to each of the operators, which will carry the information of the ordering of the operators throughout the continuation procedure. Explicitly, we choose to place the operators at the points \cite{Roberts:2014ifa}
 
 \begin{equation}
 \label{eq:inserts}
 \begin{split}
 	&z_1= e^{\frac{2\pi}{\beta}(t+i\epsilon_1)}\;\;,\;\;\;z_2= e^{\frac{2\pi}{\beta}(t+i\epsilon_2)}\;\;,\;\;\;z_3= e^{\frac{2\pi}{\beta}(x+i\epsilon_3)}\;\;,\;\;\;z_4= e^{\frac{2\pi}{\beta}(x+i\epsilon_4)}\;\;\\
 	&\bar{z}_1= e^{-\frac{2\pi}{\beta}(t+i\epsilon_1)},\;\;\,\bar{z}_2= e^{-\frac{2\pi}{\beta}(t+i\epsilon_2)},\;\;\;\bar{z}_3= e^{\frac{2\pi}{\beta}(x-i\epsilon_3)}\;\;,\;\;\,\;\bar{z}_4= e^{\frac{2\pi}{\beta}(x-i\epsilon_4)}\;\;.\\
 \end{split}
 \end{equation}\\[-4ex]

 In particular, this implies for the conformal cross ratios $z$ and $\bar{z}$ that
 
\begin{equation}
\label{eq:zcont}
z\equiv\frac{z_{12}z_{34}}{z_{13}z_{24}}\approx -e^{\frac{2\pi}{\beta}(x-t)}\epsilon_{12}^*\epsilon_{34}~~~~\text{and}~~~~~\bar{z}\equiv\frac{\bar{z}_{12}\bar{z}_{34}}{\bar{z}_{13}\bar{z}_{24}}\approx -e^{-\frac{2\pi}{\beta}(x+t)}\epsilon_{12}^*\epsilon_{34}~,
\end{equation}\\[-4ex]

 with $\epsilon_{ij}=i(e^{\frac{2\pi }{\beta}i\epsilon_i}-e^{\frac{2\pi}{\beta}i\epsilon_j})$.
 The continuation is then performed by increasing the continuation parameter $t$ to a configuration where $t>x$. As a result\footnote{As in \cite{Roberts:2014ifa}, we omit the final step of smearing the operators in real time and setting the $\epsilon_j$'s to zero.}, the operators now live on a Lorentzian sheet of the four-point function, with their ordering dictated by the hierarchy of the $\epsilon_j$'s.
 In particular, we note that the choice of $\epsilon_1<\epsilon_3<\epsilon_2<\epsilon_4$ corresponds to the out-of time order configuration introduced in equation (\ref{eq:OTOCdef}).\smallskip

\begin{figure}[h]
  \begin{centering}
  \begin{subfigure}[scale=.3]{0.32\textwidth}
  	\includegraphics[scale=.3]{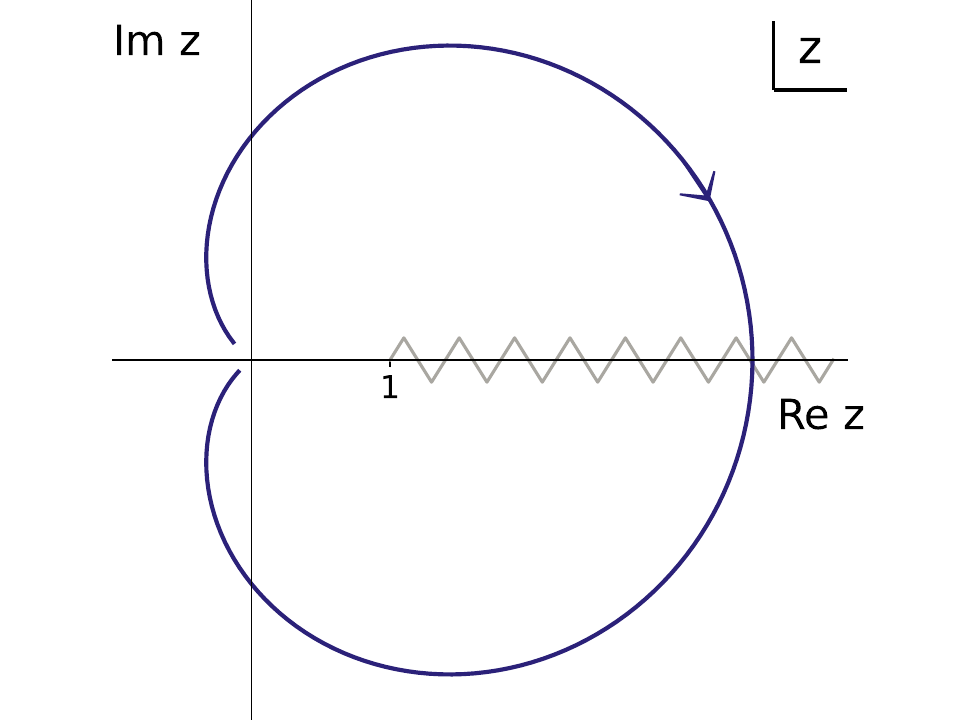}
  	\caption{}
  	\label{fig:contours_1}
  \end{subfigure}
    \begin{subfigure}[scale=.3]{0.32\textwidth}
  	\includegraphics[scale=.3]{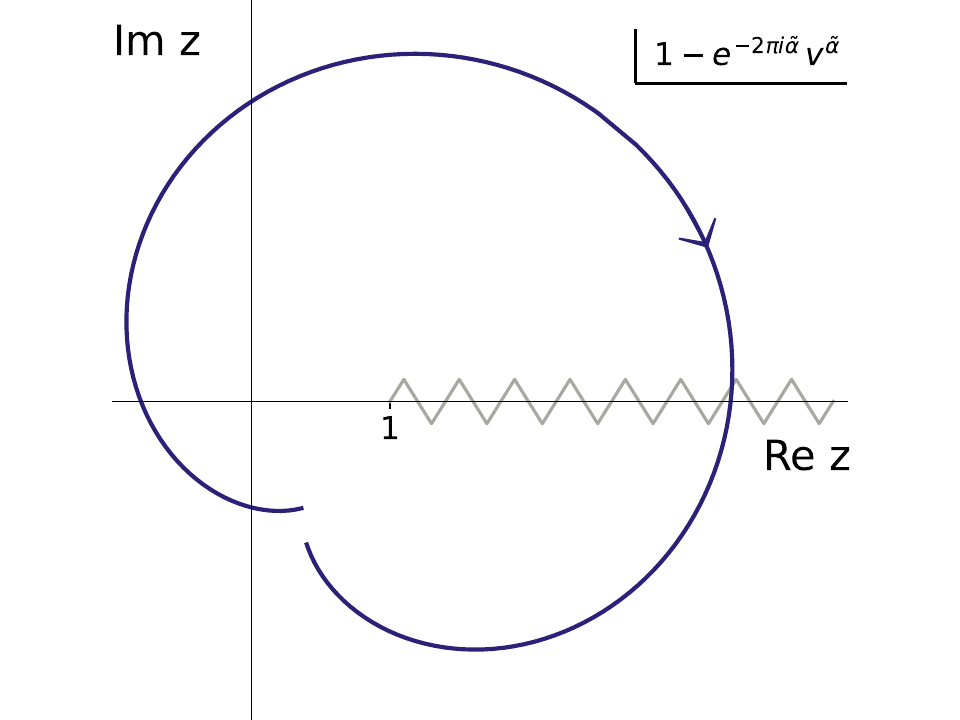}
  	\caption{}
  	\label{fig:contours_2}
  	\end{subfigure}
    \begin{subfigure}[scale=.32]{0.3\textwidth}
  	\includegraphics[scale=.3]{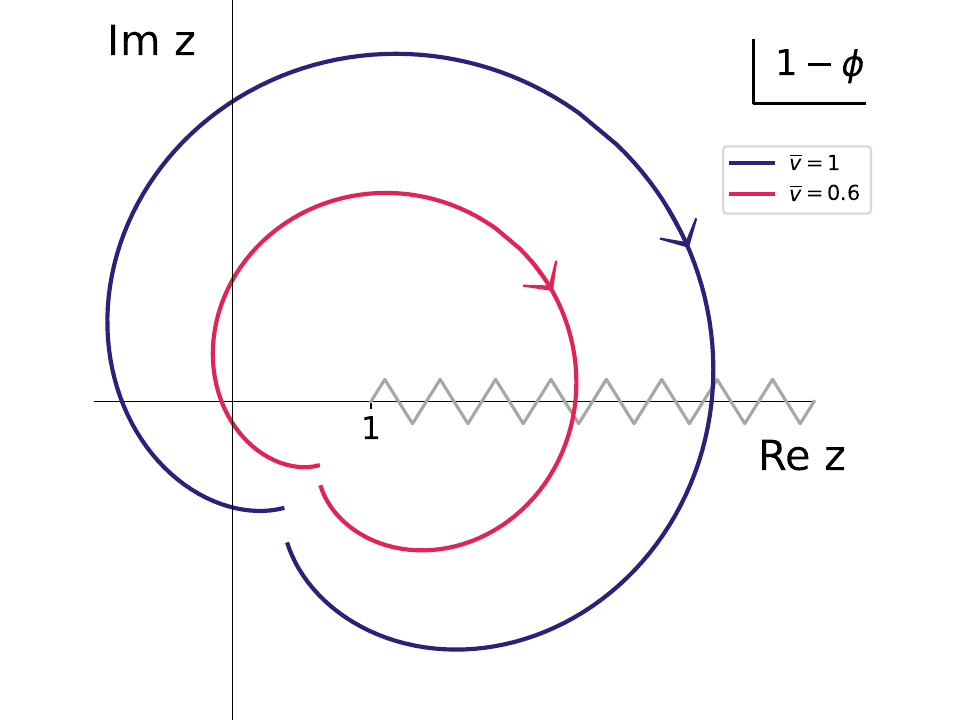}
  	\caption{}
  	\label{fig:contours_3}
  	\end{subfigure}
  	\caption{(a) The path traced out by the cross-ratio $z$ under analytic continuation to out-of-time order. (b) The contour of the argument of the conformal block $1-e^{-2\pi\,i\tilde{\alpha}}\,v^{\tilde{\alpha}}$ with $v=1-z$ under analytic continuation. (c) The contour of the argument of the compact formulation $\phi=e^{-2\pi\,i}\,v^{\tilde{\alpha}}\bar{v}^{\tilde{\alpha}}$. All contours have the common feature that they encircle the branch point at $z=1$. 
   }
    \label{fig:contours}
\end{centering}
\end{figure}

 During the continuation, the cross-ratio $z$ traces out the contour depicted in Figure \ref{fig:contours_1}. The crucial feature here is that the contour encircles the branch point $z=1$ of $f(z,\bar{z})$, therefore crosses the branch cut and moves to another sheet of the function. Concretely, this means that factors of type $(1-z)^a$ pick up a phase under continuation:
 
 \begin{equation}
 \label{eq:phase}
 	(1-z)^a \longrightarrow e^{- 2 \pi i\,a}\,(1-z)^a\;\;.
 \end{equation}\\[-4ex]
 
 Terms depending on the antiholomorphic cross-ratio $\bar{z}$ transform trivially, as the $\bar{z}$-contour does not cross any branch cuts. Having reviewed the prescription to continue Euclidean correlators to out-of time order Lorentzian ones, let us now apply it to expressions (\ref{eq:4ptcbe}) and (\ref{eq:4ptc}).

\subsection{Conformal Block Expansion}

First we derive the OTOC corresponding to the conformal block expansion of the heavy-light four point function. In this case, from (\ref{eq:4ptcbe}) we read off

\begin{equation}
\label{eq:contcbe}
f(z,\bar{z})=\sum_{h}\;C(h,h_L,h_H,c)\;z^{2h_L}\bar{z}^{2h_L}\;\left|\,\mathcal{F}(h_L,h_H,h,z)\,\right|^{2}\; \;.
\end{equation}\\[-4ex]

To continue this expression to Lorentzian times, we start with the holomorphic conformal block on the Euclidean sheet, which is given by (\ref{eq:blocklit}):

 \begin{equation}
 \label{eq:blockfact}
 	 \mathcal{F}^{\;\text{I}\,}(h_L,h_H,h,z)\; =\;\mathcal{F}^{\;\text{I}\,}_0(z)\;\left(\frac{1\!-\!(1\!-\!z)^{\tilde{\alpha}}}{\tilde{\alpha}}\right)^{h}\,{}_2F_1(h,\,h;\,2h;1\!-\!(1\!-\!z)^{\tilde{\alpha}})\;\;
 \end{equation}\\[-4ex]
In order to facilitate the analysis in the next section, we have grouped together all $h$-independent terms into the "vacuum"\footnote{Note that the vacuum conformal block does not appear in the exchange OPE but merely appears as a factor, given that $h$ is never zero due to (\ref{eq:spec}), which is a consequence of the non-renormalisability of the vacuum in Liouville theory. } conformal block

\begin{equation*}
 	\mathcal{F}^{\;\text{I}\,}_0(z)\;=\;\;(1\!-\!z)^{(\tilde{\alpha}-1)h_L}\;\left(\frac{1\!-\!(1\!-\!z)^{\tilde{\alpha}}}{\tilde{\alpha}}\right)^{-2h_L}\;\;.
\end{equation*}\\[-4ex]

When performing the continuation for (\ref{eq:contcbe}), equation (\ref{eq:phase}) now instructs us to move across the branch cut of $(1\!-\!z)^a$, and as a consequence we pick up a phase for each of the respective factors. However, we also need to be careful about the multivaluedness of the hypergeometric function itself. As $z$ moves along the contour, the argument $1\!-\!(1\!-\!z)^{\tilde{\alpha}}$ traces out the path depicted in Figure \ref{fig:contours_2} and thus crosses the branch cut of the hypergeometric function. We briefly review in Appendix \ref{sec:hypdisc} how this can be dealt with, and find that after continuation to out-of-time order the conformal block takes the form

\begin{equation}
\label{eq:contbloc}
\begin{split}
	\mathcal{F}^{\text{II}}(z)\;
 	&=\;\frac{\Gamma(2h)\,\Gamma(1\!-\!h)}{\Gamma(h)}\;\,\mathcal{F}^{\text{II}}_0(z)\;\left(\frac{1\!-\!\psi}{\tilde{\alpha}}\right)^{h}\,\Big{\{}(\uminus 1)^{h-1}\psi^{h-1}(1\!-\!\psi)^{1-2h}\;{}_2F_1(1\!-\!h,1\!-\!h;1;\psi^{\uminus 1})\\
 	 &~~~~~~~~~~~~~~~~~~~~~~~~~~~~~~~~~~~~~~~~~~~~~~~~~~~~~~~~~~~~~~~~+ e^{2\pi i \,h}(\uminus 1)^{h}\,(1\!-\!\psi)^{1-2h}\;{}_2F_1(1\!-\!h,\,1\!-\!h;\,1;\,\psi)\Big{\}}\;\;,
\end{split}
\end{equation}

where we have defined the shortcut $\psi = e^{-2\pi i\,\tilde{\alpha}}(1-z)^{\tilde{\alpha}}$, and the vacuum factor on the second sheet reads

\begin{equation}
 	\mathcal{F}_0^{\text{II}}(z)\;=\;e^{-2\pi i \, (\tilde{\alpha}-1)h_L}\;(1\!-\!z)^{(\tilde{\alpha}-1)h_L}\;\left(\frac{1\!-\!\psi}{\tilde{\alpha}}\right)^{-2h_L}\;\;.
\end{equation}\\[-4ex]

Finally, we expand (\ref{eq:contbloc}) for small $z$ and $\frac{h_H}{c}$ in order to obtain the late time limit\footnote{The limit of $\frac{h_H}{c}$ was a technical assumption in the derivation for the semi-classical heavy-light blocks in \cite{2015JHEP...11..200F}, and existing literature \cite{Roberts:2014ifa,Liu:2018iki,Chang:2018nzm} has expanded their blocks in this parameter. Nevertheless, it is anticipated that the block is valid even outside of this regime.}, which will be needed for the analysis in the next section.

\begin{equation}
\label{eq:asymp}
 	\mathcal{F}^{\text{II}}(z)\;\sim\;\,\left(-\frac{ie^{2\pi i h}}{\cos \pi h}\right)\;\frac{\Gamma(2h)\Gamma(1\!-\!h)}{\Gamma(2\!-\!2h)\,\Gamma(h)^3}\;\,\mathcal{F}^{\text{II}}_0(z)\,\left(z-\frac{24\,\pi i\,h_H}{c}\right)^{1-h}\,
\end{equation}\\[-3ex]

Note that the vacuum factor, which is usually the driving force behind maximal chaos, is only part of the story here. To make the connection with \cite{Roberts:2014ifa}, we remark that the combination $z^{2h_L}\mathcal{F}_0^{\text{II}}$ in the above limit becomes

\begin{equation}
    z^{2h_L}\mathcal{F}_0^{\text{II}}\sim\left(\frac{1}{1-\frac{24\pi i h_H}{c z}}\right)^{2h_L}.
\end{equation}\\[-4ex]

To summarise, the OTOC is given by (\ref{eq:form}) and (\ref{eq:contcbe}), where the holomorphic block is now on the second Riemann sheet (\ref{eq:contbloc}). As noted above, there is no change to the antiholomorphic factor as there are no branch cuts crossed for $\bar{z}$, i.e. $	\mathcal{F}^{\,\text{II}}(\bar{z})=\mathcal{F}^{\,\text{I}}(\bar{z})$.
\smallskip

\subsection{Compact Form}

Moving on to the compact form (\ref{eq:4ptc}) of the heavy-light four-point function, we carry out the same procedure. We now have

\begin{equation}
f(z,\bar{z}) =\;\tilde{\omega}(h_L,\tilde{\alpha},b)\;\, |z|^{4h_L}\,|1-z|^{2h_L(\tilde{\alpha}-1)}\; {}_2F_1(2h_L,\, 2h_L;\, 4h_L;\, 1-|1-z|^{2\tilde{\alpha}})\;
\end{equation}\\[-4ex]

Again, all factors of $(1\!-\!z)^a$ pick up a phase, and we need to closer examine the path traced out by the argument of the hypergeometric function upon continuation. The presence of the antiholomorphic factor has the effect of rotating and rescaling the contour, depicted in Figure \ref{fig:contours_3}, but it nevertheless remains centred around $z=1$ and the hypergeometric branch cut is inevitably crossed. We thus find 

\begin{equation}
\label{eq:OTOCclosed}
\begin{split}
    f^{\,\text{II}}(z,\bar{z})\,&=\;\hat{\omega}(h_L,\tilde{\alpha},b)\;\, |z|^{4h_L}\;|1\!-\!z|^{2(\tilde{\alpha}\uminus1)h_L}\Big{\{} \;\left(\uminus\phi\right)^{2h_L-1} (1\!-\!\phi)^{1-4h_L}\,\,{}_2F_1(1\!-\!2h_L,\,1\!-\!2h_L;1\,;\,\phi^{-1})\\
 	&~~~~~~~~~~~~~~~~~~~~~~~~~~~~~~~~~~~~~~~~~~~~~~~~~~~~~~~~~+e^{4\pi i\,h_L}(\uminus1)^{2h_L}(1-\phi)^{1-4h_L}\,{}_2F_1(1\!-\!2h_L,\,1\!-\!2h_L;1\,;\,\phi)\Big{\}},
\end{split}
\end{equation}

with $\phi = e^{-2\pi i\,\tilde{\alpha}}|1\!-\!z|^{2\tilde{\alpha}}$ and. We have redefined the constant $\hat\omega$ to absorb additional prefactors as

\begin{equation}
    \hat{\omega}(h_L,\tilde{\alpha},b)\;=\;\tilde{\omega}(h_L,\tilde{\alpha},b)\;\,e^{-2\pi i (\tilde{\alpha}-1)h_L}\;\frac{\Gamma(4h_L)\,\Gamma(1\!-\!2h_L)}{\Gamma(2h_L)}\;.
\end{equation}\\[-4ex]

Finally, let us take the limit of small $z$ and $\frac{h_H}{c}$, and for $h_L\gg 1$ at large $c$ we find up to constant factors

\begin{equation}
\label{eq:OTOCclosedapprox}
    f^{\,\text{II}}(z,\bar{z})\;\sim\;|z|^{4h_L}\;
    \left(z+\bar{z} - \frac{24 \pi i \,h_H}{c}\right)^{-4h_L}\;\;.
\end{equation}\\[-4ex]

In summary, the compact form of the OTOC is given by (\ref{eq:form}) with the function $f(z,\bar{z})$ continued to the second sheet (\ref{eq:OTOCclosed}). As we shall see in the next section, the two versions of the OTOC will offer complementary insights into quantum chaos in Liouville theory.\bigskip

\section{Scrambling and Quantum Chaos}
\label{sec:chaos}
Equipped with the heavy-light OTOC in closed form (\ref{eq:OTOCclosed}), and the rearranged expression in terms of a conformal block expansion (\ref{eq:contbloc}), we are ready to explore the chaotic character of Liouville theory.
The former result is ideally suited to study universal properties of the OTOC, whereas the latter sheds light on the process of scrambling and the relative importance of the conformal blocks to the four point function. In the following , we are interested in studying the behaviour of the \textit{normalised} OTOC \cite{Roberts:2014ifa}

\begin{equation}
f(z,\bar{z})=\frac{~~\big\langle H(t+i\epsilon_1)L(i\epsilon_3)H(t+i\epsilon_2)L(i\epsilon_4)\big\rangle_{\beta}}{\big\langle H(i\epsilon_1)H(i\epsilon_2)\big\rangle_{\beta}\;\big\langle L(i\epsilon_3)L(i\epsilon_4)\big\rangle_{\beta}~~~~~}\;\equiv\;\langle \text{HHLL}\rangle_{\text{OTOC}}~.
\end{equation} 
\subsection{Scrambling Time and Lyapunov Exponent}

Let us start with some of the "global" aspects of the OTOC. An important signature of chaos is the exponential decrease of the out-of time ordered four point function at scrambling time. In order to study this phenomenon, let us make use of the late-time approximation of the heavy-light correlator (\ref{eq:OTOCclosedapprox}):

 \begin{equation}
 	\langle \text{HHLL}\rangle_{\text{OTOC}}\;\sim\;|z|^{4h_L} \left(z+\bar{z}-\frac{24 \pi i\, h_H}{c}\right)^{-4h_L} 
 \end{equation}\\[-4ex]

In the above expression, we have omitted all constant factors as they are unimportant for the following analysis. Let us restore the explicit time dependence using (\ref{eq:zcont}), which furthermore justifies the approximation $z\sim\bar{z}$ at late times. We obtain

 \begin{equation}
 \label{eq:decayotoc}
 	\langle \text{HHLL}\rangle_{\text{OTOC}} \; \sim \; \left(1-\frac{12\pi i\,h_H}{c\,z}\right)^{-4h_L}\;=\;\,\left(1+\frac{12\pi i}{\epsilon_{12}^*\epsilon_{34}} \,e^{\frac{2\pi}{\beta}(t-x-t_*)}\right)^{-4h_L}\;\;.
 \end{equation}\\[-4ex]

Note that $h_L$ scales with the central charge, so the exponent is large and negative. As time progresses, the exponential term in the bracket grows larger, and the overall OTOC decays as expected. We have introduced a timescale $t_*$ governing the decay of the OTOC, called "scrambling time":

\begin{equation}
 \label{eq:ts}
 	t_{*} := \frac{\beta}{2\pi} \log\frac{c}{h_H}\;\;.
 \end{equation}\\[-4ex]
 
The essential insight here is the logarithmic dependence on system size, which was anticipated in \cite{Sekino:2008he}\cite{Shenker:2013pqa} and the logarithmic scaling in the central charge was confirmed at the level of the identity block approximation in \cite{Roberts:2014ifa}. Here we have extended this observation to the full heavy-light OTOC in semiclassical Liouville theory, for which it was possible to resum the conformal block expansion thanks to out technical results in Section \ref{sec:cbe}. 
We should remark that in claiming logarithmic dependence on the central charge, we are once again stretching the formalism beyond the original range of validity for its parameters, as in \cite{Roberts:2014ifa}. Initially, we assumed in (\ref{eq:weights}) that $h_H/c$ is kept constant in the semiclassical limit. In order to derive the late time approximation (\ref{eq:OTOCclosedapprox}), we have moreover assumed that this ratio is small. Our claim of logarithmic scaling of $t_*$ with the central charge in (\ref{eq:ts}) however is only meaningful if $h_H$ is kept fixed independently of $c$, which a priori lies outside our allowed parameter range. In the identity block analysis in \cite{Roberts:2014ifa} this step was justified by the fact that matching to the gravity calculation suggests that the conformal blocks (\ref{eq:blocklit}) are valid for a wider range of parameters, and we shall take the same stance here. Finally, from (\ref{eq:decayotoc}) we can read off the Lyapunov exponent from the exponential factor:

\begin{equation}
    \lambda_L=\frac{2\pi}{\beta}
\end{equation}\\[-4ex]

Therefore, semiclassical Liouville theory saturates the bound on chaos in \cite{Maldacena:2015waa}. Once again we have demonstrated this result at the level of the full OTOC, improving on attempts in the literature arguing on a block by block basis. Even though the individual blocks still decay at the same rate \cite{Liu:2018iki}\cite{Hampapura:2018otw}\cite{Chang:2018nzm}\cite{Kusuki:2019gjs}, as can be easily read off from (\ref{eq:asymp}), there is no guarantee that the sum over all conformal blocks should follow the same behaviour, and for Liouville CFT we have closed this gap with the above result. Having derived a compact form for the OTOC, we were therefore able to estalish that Liouville theory is maximally chaotic, as was suggested from the lattice Liouville CFT construction in \cite{Turiaci:2016cvo}. 

\subsection{Scrambling and Conformal Block Dominance}

Having extracted these two characteristics from our compact expression for the OTOC, let us also see what we can learn from the conformal block expansion (\ref{eq:contcbe}). Our derivation in Chapter \ref{sec:HL4} has provided us with a list of conformal dimensions (\ref{eq:spec}) of all the operators exchanged in the semiclassical theory, and moreover we have knowledge of the conformal blocks (\ref{eq:contbloc}) and OPE coefficients (\ref{eq:OPEs}), giving us full control over the individual contributions to the OTOC. Equipped with this information, our goal is to investigate how the relative importance of the individual blocks is affected by scrambling. Before we address this question for the out-of-time ordered correlator, let us first build some intuition based on the Euclidean four-point function.
In the Euclidean theory, the blocks exchanged by the conformal block expansion (\ref{eq:4ptcbe}) are of the form

\begin{equation}
z^{2h_L}\bar{z}^{2h_L}\mathcal{F}^{\;\text{I}}(h_L,h_H,h,z)\,\mathcal{F}^{\;\text{I}}(h_L,h_H,h,\bar{z})\;\;,
\end{equation}\\[-4ex]

with $\mathcal{F}^{\;\text{I}}(z)$ given by (\ref{eq:blocklit}). To estimate the contribution of the block, let us once again expand in $z$ and $\bar{z}$ respectively to obtain the late-time limit. We find that the block takes on a very simple form:

\begin{equation}
    |z|^{4h_L}\;|\mathcal{F}^{\;\text{I}}(h_L,h_H,h,z)\,|^2\;\sim\;|z|^{2h}
\end{equation}\\[-4ex]

Recalling that the exchanged states in the semiclassical Liouville four-point function are given by $h=2h_L+r$ for $r\in \mathbb{N}_0$, we can write down the ratio of the lowest weight block, given by $r=0$, to a generic block in this limit:

\begin{equation}
    \frac{|\,\mathcal{F}_{\;r=0}^{\;\text{I}}\;(z)\,|^2}{|\,\mathcal{F}_{\;r\neq 0}^{\;\text{I}}\;(z)\,|^2}\;\sim\;\frac{1}{\;|z|^{2r}}
\end{equation}\\[-4ex]

Given that z is small, we see that on the Euclidean sheet the $r=0$ block dominates over all the other blocks running in the four-point function at late times. This is echoing the expectation that in a generic CFT with large central charge, the identity block dominates the exchange \cite{Hartman:2013mia}, at least in the Euclidean theory. Because there is no identity block in Liouville theory, this role is instead taken on by the next lowest-weight state.\medskip

The story changes drastically for the out-of-time ordered correlator. If we instead investigate block dominance on the second Riemann sheet, where the holomorphic conformal blocks in the late time limit are given by (\ref{eq:asymp}), we find that each term in the conformal block expansion is of the form

\begin{equation}
\label{eq:fullblockII}
      |z|^{4h_L}\;|\mathcal{F}^{\;\text{II}}(h_L,h_H,h,z)\,|^2\;\sim\;
     c(h)\; |z|^{4h_L}\;\mathcal{F}^{\text{II}}_0(z)\,\left(z-\varepsilon\right)^{1-h}\;\bar{z}^{h}\;\;.
\end{equation}\\[-4ex]

We have defined the parameter $\varepsilon=\frac{24\,\pi i\,h_H}{c}$, proportional to the expansion parameter $\frac{h_H}{c}$, and the coefficient $c(h)$ groups together $z$-independent parameters as

\begin{equation}
c(h)\;=\;-\frac{i e^{2\pi i h}}{\cos \pi h}\;\frac{\Gamma(2h)\Gamma(1\!-\!h)}{\Gamma(2\!-\!2h)\,\Gamma(h)^3}\;\;.
\end{equation}\\[-4ex]

The coefficient can be estimated using Stirling's approximation for the Gamma functions, and we find that it is of order $|c(h)|\;\sim\; 4^{2h-1}$.
Hence, we can simplify the expression to 

\begin{equation}
\label{eq:fullblockIIshort}
      |z|^{4h_L}\;|\mathcal{F}^{\;\text{II}}(h_L,h_H,h,z)\,|^2\;\sim\;
      \;4^{2h}\,|z|^{4h_L}\;\mathcal{F}^{\text{II}}_0(z)\;\left(1-\frac{\varepsilon}{z}\,\right)^{-h}\;\;.
\end{equation}\\[-4ex]

We have additionally made use of the fact that $z\approx\bar{z}$ at late times, and moreover that $1-h\approx -h$ in the semiclassical limit. 
With this expression at hand, it is easy to write down the ratio of the lowest-weight OTO block to a generic one. Invoking $h=2h_L+r$, we obtain

\begin{equation}
\label{eq:OTOrat}
\frac{|\,\mathcal{F}_{\;r=0}^{\;\text{II}}\;(z)\,|^2}{|\,\mathcal{F}_{\;r\neq 0}^{\;\text{II}}\;(z)\,|^2}\;\sim\; 4^{-2r}\;\left(1-\frac{\varepsilon}{z}\,\right)^{r}\;\;.
\end{equation}\\[-4ex]

The picture is now very different from the Euclidean theory. We start off our analysis by considering the early time\footnote{Strictly speaking, equation (\ref{eq:OTOrat}) is only valid in the small-$z$ limit. Nevertheless, it turns out to be qualitatively accurate and in agreement with the behaviour of the exact block in (\ref{eq:contbloc}) even at early times.} behaviour of the ratio (\ref{eq:OTOrat}), which is governed by $|z|\sim1$. In this limit, the term in brackets has an absolute value close to and slightly larger than unity. This term however is by far outweighed by the first factor $4^{-2r}$, meaning that higher-weight blocks get larger and larger with growing $r$, and consequently there is no single operator dominating the expansion. This is quite contrary to the intuition from the Euclidean theory presented above. Nevertheless, this behaviour has also been observed in the studies of \cite{Hampapura:2018otw} \cite{Chang:2018nzm}\footnote{The same result also follows from the analysis in \cite{Liu:2018iki}, after taking into account the anti-holomorphic factor of the block.}, and Liouville theory is no exception here.

The tables turn however as we approach scrambling time (\ref{eq:ts}), where the above arrangement is going to get reshuffled. At $t=t_*$, the term $\varepsilon/z$  becomes of order unity and the bracket is able to compete with the prefactor.
Once the bracket grows to $|1-\varepsilon/z|\sim 4^2$, the expression (\ref{eq:OTOrat}) increases rapidly, signalling that the lower-$r$ blocks take over the correlator.
From this point on, the configuration is dominated by the $r\!=\!0$ contribution. Compared to early times, the importance of the conformal blocks in the expansion has now been inverted. Such a reshuffling of states is special to semi-classical Liouville theory, due to the absence of any spin operators in the spectrum (\ref{eq:spec}), which in generic CFTs are known to dominate the late time OTOC \cite{Hampapura:2018otw} \cite{Chang:2018nzm}. Algebraically, we have used the condition $h=\bar{h}$ to obtain expression (\ref{eq:OTOrat}), which we proved for semiclassical Liouville theory in Chapter \ref{sec:HL4}. It would be tempting to attribute this reorganisation of dominance of conformal blocks to a Stokes phenomenon of the saddles in the path integral formulation (\ref{eq:4pt}) of the four-point function, however this is not the case due to the non-local nature of the map between path integral saddles and conformal blocks.\bigskip

\section{Discussion}
\label{sec:disc}

In this section, we want to reflect on the insights obtained in this paper, and place them in the context of holography. In the example of Liouville theory, we have provided a prescription how the conformal block expansion can be resummed to a closed form, and applied this insight to extract a scrambling time and a Lyapunov exponent from the OTOC. Given that Liouville CFT is thought to be intimately related to quantum gravity -- although perhaps not as a direct holographic dual, but rather in an ensemble sense \cite{Chandra:2022bqq, Belin:2023efa,Collier:2023fwi,Collier:2024mgv} -- 
let us comment on other proposals in which Liouville CFT might enter into the gravity picture, and explain how our results add to existing understanding.

The first is based on an observation in \cite{Jackson:2014nla}, where the authors propose to identify Liouville theory as a universal subsector appearing in any irrational CFT with a large central charge that admits a holographic dual, by matching R-matrices and exchange algebras of 2+1 D gravitational shock waves in the bulk and their dual 2D CFT. 
This argument was further developed in \cite{Turiaci:2016cvo}, where it is shown that the bulk entanglement entropy can be used to derive the stress tensor, which precisely is of the form of that of Liouville theory. The paper \cite{Vos:2020clx} reaches a similar conclusion, by demonstrating that the identity block in a mixed heavy and light correlator in a large-c CFT is captured by a Liouville correlation function. If we follow this line of thinking, Liouville theory is intimately linked to the graviton mode in semi-classical AdS$_3$, and therefore has to replicate the chaotic dynamics associated with black holes \cite{Sekino:2008he}. In our analysis we have proven that the semi-classical heavy-light correlator saturates the bound of chaos derived in \cite{Maldacena:2015waa}, providing further evidence in support of the above hypothesis, although we have explained that the mechanism in which it does so is somewhat subtle.

A second intriguing point of view of the role of Liouville theory in holography was recently suggested in \cite{Chandra:2022bqq, Belin:2023efa, Collier:2023fwi}. The authors propose that semiclassical gravity in AdS$_3$ should be dual to an average of conformal field theories, rather than any single CFT. This is well motivated from the expectation that a semiclassical observer doesn't have access to UV degrees of freedom, and is therefore perceiving a coarse grained version of an underlying microscopic theory. Moreover, this proposal is  analogous to what happens in $AdS_2$, which was shown to be dual to an ensemble of random matrix models \cite{Saad:2019lba}. In the context of the averaged-CFT proposal in \cite{Chandra:2022bqq, Belin:2023efa}, the action of a 2-boundary wormhole was shown to correspond to two copies of Liouville theory.
By demonstrating that semiclassical Liouville CFT is maximally chaotic, we also feed into this idea, where in this scenario our result is analogous to the fact that the SYK model is maximally chaotic in AdS$_2$ holography. It is interstring to note, however, that the proposal of \cite{Chandra:2022bqq, Belin:2023efa} concerns (an average of) large$-c$, large-gap CFT$_2$ which do include the identify, and the maximal scrambling exponent in these types of theories should be traced back to the identity block in the s-channel, or via channel duality to the Liouville continuum above $\frac{c-1}{24}$ in the t-channel. The precise role of the ``dressed-scramblon" mechanism uncovered in this paper for the proposals of \cite{Chandra:2022bqq, Belin:2023efa} remains to be elucidated. \medskip

Another interesting feature our analysis establishes is the reshuffling of the conformal block expansion at the scrambling time. This sheds light on the mechanism of scrambling, and reveals some of the structural changes the OTOC undergoes such that it begins to decay exponentially at scrambling time. While no single block dominates the correlator at early times, after scrambling the dominance is shifted in favour of the lowest-weight block in the spectrum. This phenomenon is special in Liouville theory due to the absence of spin-operators in the semiclassical theory.

 From a technical perspective, the relationship between path integral saddles and conformal blocks derived in Section \ref{sec:HL4} is very interesting, and to our knowledge the only example where such a map was explicitly realised. It would be interesting to see how this relation is implemented in other CFTs. A neat feature of this calculation is that in our example, the conformal block expansion nicely resums into a compact form, which is possible thanks to the discreteness of the spectrum and the evenly spaced operator dimensions. In principle, such a resummation should be possible in any CFT with these characteristics.
 Finally, it would be natural to generalise our analysis to the case of four heavy operator insertions, or indeed of higher-point OTOCs in a suitable limit \cite{Anous:2020vtw}.

\subsection*{Acknowledgements}
We thank Anton Alexeev, Tarek Anous, Tom Hartman, Luca Iliesiu, Pranjal Nayak, Eric Perlmutter and Herman Verlinde for interesting discussions and additionally Tarek Anous for comments on a draft of this work. This work has been supported in part by the Fonds National Suisse de la Recherche Scientifique (Schweizerischer Nationalfonds zur Förderung der wissenschaftlichen Forschung) through ProjectGrants200020 182513, the NCCR51NF40-141869 The Mathematics of Physics (SwissMAP), and in part by a UK Engineering and Physical Sciences Research Council (EPSRC) studentship. This work was performed in part at Aspen Center for Physics, which is supported by National Science Foundation grant PHY-2210452.

\appendix

\section[Integration Trick]{Integration Trick for $\tilde{S}_{LV}$}
\label{sec:inttrick}

For the sake of completeness, we give a brief summary of the integration trick \cite{Zamolodchikov:1995aa,2011JHEP...12..071H} needed to compute the modified Liouville action $\tilde{S}_{LV}$ in the presence of heavy operators as stated in section \ref{sec:HL4}. The key idea will be that near the singular points $z_i$ of the field configuration, the function $\frac{\partial}{\partial\eta_i}\tilde{S}_{LV}$ is constant, where $\eta_i$ is parametrising the weight of the heavy insertions. On solutions $\phi_c$ of the semi-classical equations of motion, the action can be obtained by simple integration

\begin{equation}
\label{eq:inttrick}
\tilde{S}_{LV}[\phi_c]= \int  \frac{\partial \tilde{S}_{LV}}{\partial \eta_i}d\eta_i \;,
\end{equation}\\[-4ex]

which by continuity defines $\tilde{S}_{LV}$ everywhere in the complex plane. In this section we expand on the above argument and derive the modified action in the special case of two heavy operators of equal weight, as advertised in (\ref{eq:modS}).\bigskip

Naïvely, one would go about constructing a correlation function by inserting vertex operators $V_{\alpha}=e^{2\alpha\phi}$ into the path integral built from the semi-classical action:

\begin{equation}
\label{eq:PI}
    \left\langle V_{\alpha_1}(z_1,\bar{z}_1)\dots V_{\alpha_n}(z_n,\bar{z}_n)\right\rangle=\int \mathcal{D}\phi_c\; e^{-S_{LV}[\phi_c]}\,\prod_{i=1}^n e^{\frac{\alpha_i}{b}\phi_c}~,
\end{equation}\\[-4ex]

with the action given by (\ref{eq:SLV})

\begin{equation*}
	b^2\;S_{LV}\;=\;\frac{1}{16\pi}\int_{D} d^2\xi\;\left\{\left(\partial_a\phi_c\right)^2+16\pi\mu b^2 e^{\phi_c}\right\} \;+\; \frac{1}{2\pi}\oint_{\partial D}\;\phi_c d\theta + 2\log R + \mathcal{O}(b^2)\;\;.
\end{equation*}\\[-4ex]

This naïve construction breaks down however whenever the $\phi_c$-integration hits a field configuration where heavy operators are present. In this case, the above action diverges and has to be appropriately regularised. In \cite{Zamolodchikov:1995aa} it was shown that this can be done by removing small disks $d_i$ of radius $\epsilon$ around the heavy insertion points from the domain $D$ in the spacetime integral of $S_{LV}$, while renormalising the heavy vertex operators as 

\begin{equation}
V_{\frac{\eta_i}{b}}(z_i,\bar{z}_i)\equiv \epsilon^{\frac{2\eta_i^2}{b^2}} \exp \left(\frac{\eta_i}{2\pi}\oint_{\partial d_i}\phi_c d\theta\right),
\end{equation}\\[-3ex]

Both effects can be absorbed into a modified action $\tilde{S}_{LV}$, which now remains finite in the limit of $\epsilon \to 0$:\\[.5ex]
\begin{equation}
\begin{split}
\label{eq:modS2}
   b^2 \tilde{S}_{LV}&=\frac{1}{16\pi}\int_{D-\cup_i d_i} d^2\xi\;\left\{\left(\partial_a\phi_c\right)^2+16\pi\mu b^2 e^{\phi_c}\right\} \;+\; \frac{1}{2\pi}\oint_{\partial D}\;\phi_c d\theta + 2\log R\\
   &~~~\;\; - \sum_{i} \left(\frac{\eta_i}{2\pi}\oint_{\partial d_i}\phi_c d\theta+ 2\eta_i^2  \log\epsilon\right) 
\end{split}
\end{equation}\\[-4ex]

With this issue resolved, let us return to the construction (\ref{eq:PI}) of correlators, with $\tilde{S}_{LV}$ replacing $S_{LV}$ and the product in $i$ only running over light operators. We wish to evaluate the path integral by means of a saddle point approximation, where the saddles $\phi_{\eta_i}$ are solutions to the semi-classical equations of motion (\ref{eq:eom}). As explained in \cite{2011JHEP...12..071H}, the presence of heavy operators modifies the equations of motion by adding delta functions corresponding to the heavy insertions, which in turn alters the solutions $\phi_{\eta_i}$ as indicated in subscript.  In the following, we treat the special case of two identical heavy operators $\alpha_1=\alpha_2=\eta_H/b$, in which case the equation of motion becomes

\begin{equation}
\label{eq:eom2}
\partial\bar{\partial}\phi_c = 2\lambda e^{\phi_c} - 2\pi \eta_H\,\delta^2(\xi-\xi_1) - 2\pi \eta_H\,\delta^2\left(\xi-\xi_2\right)~,
\end{equation}\\[-4ex]

with $\lambda=\pi\mu b^2$. Near the singular points $\xi_1$ and $\xi_2$ the exponential term becomes negligible \cite{2011JHEP...12..071H} as long as they Seiberg bound $Re(\eta_H)<\frac{1}{2}$ is obeyed, and a solution is given by 
\begin{equation}
\label{eq:nearz}
\phi_{\eta_H}=-4\eta_H\log|z-z_i| +C_i~~~~\text{as}~~z\to z_i~~,~\text{with}~~C_i\sim\mathcal{O}(1). 
\end{equation}\\[-4ex]

A full solution to (\ref{eq:eom2}) with the correct behaviour (\ref{eq:nearz}) near the insertion points was found to be \cite{2011JHEP...12..071H}
\begin{equation}
\label{eq:phiapp}
 	e^{\phi_{\eta_H}} \; = \; \frac{1}{\lambda} \frac{\kappa^2}{\;\left(\;\kappa^2\, |z-z_1|^{2\eta_H}\, |z-z_2|^{2-2\eta_H} - \frac{1}{(1-2\eta_H)^2\,|z_{12}|^2}\,|z-z_1|^{2-2\eta_H}\,|z-z_2|^{2\eta_H}\;\right)^2}\;\;.
 \end{equation}\\[-4ex]

Now to the crucial step: observe that the partial derivative of $\tilde{S}_{LV}$ with respect to $\eta_H$ in the limit $\epsilon\to 0$ evaluates to a constant

\begin{equation}
\label{eq:parteta}
    \lim_{\epsilon\to 0}\;\frac{\partial}{\partial \,\eta_H}\tilde{S}_{LV} =  \; -\lim_{\epsilon\to 0} \;\sum_i \left(\frac{1}{2\pi}\oint_{d_i}\phi_c\, d\theta+ 4\eta_i  \log\epsilon\right) = -C_1 - C_2~,
\end{equation}\\[-4ex]

where we have used the form (\ref{eq:nearz}) for the field $\phi_c$ near the insertion point. As we are only interested in evaluating the action $\tilde{S}_{LV}$ on a solution $\phi_c$ of the semi-classical equations of motion (where the variation of $\tilde{S}_{LV}$ with respect to the field vanishes), this is enough to determine $\tilde{S}_{LV}$ by simply integrating (\ref{eq:parteta}). All that is left for us to do is to determine the constants $C_1$ and $C_2$, which can be done as follows. First, take the logarithm of the full solution (\ref{eq:phiapp})
\begin{equation}
\begin{split}
\label{eq:philog}
    \phi_{\eta_H}&= 2\pi i N - \log\lambda + 2\log \kappa \\&
    ~~- 2\log\Big(\kappa^2\, |z-z_1|^{2\eta_H}\, |z-z_2|^{2-2\eta_H} - \frac{1}{(1-2\eta_H)^2\,|z_{12}|^2}\,|z-z_1|^{2-2\eta_H}\,|z-z_2|^{2\eta_H}\Big)~~,
\end{split}
\end{equation}

where the integer $N$ labels the branch of the logarithm. Next, we expand around the singular points $z_1$ and $z_2$ to extract the constants in (\ref{eq:nearz}). Assuming $\eta_H<\frac{1}{2}$, we find

\begin{equation}
\begin{split}
C_1&=2\pi i N-\log \lambda-2\log \kappa - (4-4\eta_H)\log |z_{12}| \\
C_2&=2\pi i (N-1)-\log\lambda + 2\log\kappa + 4\log(1-2\eta_H)+4\eta_H\log|z_{12}|\\
\end{split}
\end{equation}\\[-4ex]

Now we have everything needed to integrate (\ref{eq:parteta}) in $\eta_H$. The integration constant $C_{int}$ can be fixed by observing that the field (\ref{eq:philog}) evaluated at $\eta_H=0$ lies in the $SL(2,\mathbb{C})$-orbit\footnote{The orbit is parametrised by $\phi_c=2\pi i N + \pi i -\log \lambda-2\log\left(|az+b|^2 + |cz+d|^2\right)$, with $a,b,c,d\in\mathbb{C}$ such that $ad-bc=1$. It is convenient to relabel $\kappa=i\tilde\kappa$ to eliminate the relative sign in the logarithm.} of minus the round sphere \cite{2011JHEP...12..071H}
\begin{equation}
    \phi_0=\pi i + 2\pi i N - \log \lambda -2\log \left(1+z\bar{z}\right).
\end{equation}

For this special case, the action (\ref{eq:modS2}) can be evaluated directly, which fixes the $\eta_H$-independent integration constant
\begin{equation}
\label{eq:intcons}
C_{int}=b^2\tilde{S}_{LV}[\phi_0]=2\pi i N + \pi i -\log \lambda -2 ~.
\end{equation}

Finally we integrate (\ref{eq:parteta}) using (\ref{eq:intcons}), resulting in the final expression for the modified action given in (\ref{eq:modS}):
 \begin{equation}
 \begin{split}
 	b^2\tilde{S}_{LV}\; &= \; 2 \pi i \left(N + 1/2\right) (1 - 2\eta_H) + (2\eta_H -1)\log\lambda + 4(\eta_H-\eta_H^2) \log |z_{12}|\\
 		&\;\;+2\left[(1-2\eta_H) \log (1-2\eta_H)-(1-2\eta_H) \right]
 \end{split}
 \end{equation}\\[-4ex]

\section{Moduli Space Integral}
\label{sec:int}

In this section, we fill in some of the gaps skipped over in the evaluation of the continuous moduli space integral as a hypergeometric function. The starting point is the integral\footnote{Since $\sigma_L\in\mathbb{R}$, the numerator is a multi-valued function for negative values of $\kappa$. However, we shall see that this ambiguity only modifies the overall coefficients, which plays no role in our analysis.}

 \begin{equation}
 	\label{eq:modintap}
  	\mathcal{H}\;\equiv\;\int_{-\infty}^{\;\infty} d\kappa \; \frac{\;\kappa^{4\sigma_L +1 }}{(\zeta_3\kappa^2- \chi_3 \;{+ \;i \varepsilon })^{2\sigma_L}(\zeta_4\kappa^2- \chi_4\;{+ \;i \epsilon})^{2\sigma_L}}\;\;.
 \end{equation}\\[-4ex]

First, we notice that the integrand has an even denominator, so we split the integral at zero and factor out a phase from the part along the negative real axis:

 \begin{equation}
 	\label{eq:modint2}
	\frac{1+(e^{\pi i+2\pi i M})^{4\sigma_L+1}}{(\zeta_3\zeta_4)^{2 \sigma_L}}\;\; \textcolor{gray}{\textcolor{black}{\int_{0}^{\infty} d\varkappa \; \frac{\varkappa^{4 \sigma +1}}{(\varkappa^2- \chi_3/\zeta_3 \;+ \;i \varepsilon )^{2\sigma_L}(\varkappa^2- \chi_4/\zeta_4\;+ \;i \varepsilon)^{2\sigma_L}}}}\;\;.
 \end{equation}\\[-4ex]

Here, $M\in\mathbb{Z}$ reflects the choice of branch for the multi-valued integrand. In order to keep the notation simple let us ignore the prefactor for now. Next, we make the substitution $y=\varkappa^2$, obtaining 

\begin{equation}
    \int_{0}^{\infty} \frac{dy}{2}\; \frac{y^{2\sigma_L}}{(y- \chi_3/\zeta_3 \;+ \;i \varepsilon )^{2\sigma_L}(y- \chi_4/\zeta_4\;+ \;i \varepsilon)^{2\sigma_L}}\\[2ex]
\end{equation}\\[-4ex]

Introducing Feynman parameters $x_1$ and $x_2$, we can rewrite the expression as

\begin{equation}
\label{eq:feyn}
	\frac{\Gamma(4\sigma_L)}{\Gamma(2\sigma_L)^2}\;\int_{0}^{\infty} dy \int_{0}^1 d x_1  \int_0^1dx_2 \; \delta(x_1+x_2-1)\,\; \frac{(x_1 x_2)^{2\sigma_L-1}y^{2\sigma_L}}{[x_1(y- \chi_3/\zeta_3 \;+ \;i \varepsilon )+x_2(y- \chi_4/\zeta_4 \;+ \;i \varepsilon )]^{4\sigma_L}}\;.
\end{equation}\\[-4ex]

After a final step of algebra, we arrive at the form

 \begin{equation}
 	\label{eq:zint}
 	\frac{\Gamma(4\sigma_L)}{\Gamma(2\sigma_L)^2}\;\int_0^{1}  \,d x_1 \int_0^1 d x_2\;\frac{(x_1 x_2)^{2\sigma_L-1}}{(x_1+x_2)^{4\sigma_L}}\;\; \delta(x_1+x_2-1)\,\int_{0}^{\infty} dy\;\frac{y^{2\sigma_L}}{\left[y-{\Xi}\,\right]^{4\sigma_L}}\;\;,\smallskip
 \end{equation}\smallskip

where we have grouped together $y$-independent terms into

 \begin{equation}
 	\Xi\;=\; \frac{x_1 \frac{\chi_3}{\zeta_3}+x_2\frac{\chi_4}{\zeta_4}}{x_1+x_2} + i \varepsilon \;\;.
 \end{equation}\smallskip

With (\ref{eq:zint}) we have finally obtained a form where we can straight-forwardly perform the $y$-integration. In order to do so, we recognise that it is simply a special case of Euler's integral representation of the hypergeometric function, which reads

 \begin{equation}
 \label{eq:euler_1}
 	{}_2F_1(a,\,b\,;\,c;\,z)\;=\;\frac{\Gamma(c)}{\Gamma(b)\Gamma(c-b)}\;\int_0^1\;dt\; t^{b-1}(1-t)^{c-b-1}(1-z t)^{-a}\;\;.
 \end{equation}\smallskip

 For this formula to apply to our integral in (\ref{eq:zint}), we need to extend the integration contour to the positive half of the real line. By substituting $y=\frac{t}{1-t}$, the domain is mapped to $[0,\infty)$, and we recover an alternative form of the integral representation

 \begin{equation}
     \label{eq:euler_2}
     {}_2F_1\big(a,\,b;\,c;\,1-\frac{1}{u}\big)=\frac{\Gamma(c)}{\Gamma(b)\Gamma(c-b)}\;u^{a}\,\int_0^{\infty}\,dy\;\, \frac{y^{b-1}}{(1+y)^{a-c}(u+y)^a},
 \end{equation}\smallskip

 where we have introduced $u\equiv1/(1-z)$. Comparing with (\ref{eq:zint}), we find that the $y$-integral evaluates to

 \begin{equation}
 \begin{split}
 	\int_0^{\infty} dy\;\frac{y^{2\sigma_L}}{[y-\Xi]^{4\sigma_L}}\;&=\;\left(-\Xi\right)^{-4\sigma_L}\frac{\Gamma(2\sigma_L-1)\,\Gamma(2\sigma_L+1)}{\Gamma(4\sigma_L)}\;{}_2F_1\Big(4\sigma_L,\, 2\sigma_L+1;\, 4\sigma_L;\, 1+\frac{1}{\Xi}\Big)\\[1ex]
  &=\;\left(-\Xi\right)^{1-2\sigma_L}\frac{\Gamma(2\sigma_L-1)\,\Gamma(2\sigma_L+1)}{\Gamma(4\sigma_L)}
 \end{split}\;\;.
 \end{equation}\smallskip

  All that is left for us to do now is to evaluate the integrals over the two Feynman parameters $x_1$ and $x_2$. The first integral is trivial due to the delta function, and the second integration can once again be  performed by comparison with the hypergeometric function (\ref{eq:euler_1}). Collecting all prefactors, we obtain

 \begin{equation}
 	\mathcal{H}\;=\; \frac{1+(e^{\pi i+2\pi i M})^{4\sigma_L+1}}{2\,(\zeta_3\zeta_4)^{2\sigma_L}}\;\frac{\Gamma(2\sigma_L+1)\Gamma(2\sigma_L-1)}{\Gamma(4\sigma_L)}\left(-\frac{\zeta_4}{\chi_4}\right)^{2\sigma_L-1}{}_2F_1\big(2\sigma_L-1,\, 2\sigma_L;\, 4\sigma_L;\, 1-\frac{\chi_3\zeta_4}{\zeta_3\chi_4}\big)\;.
 \end{equation}\\[-3ex]

In order to obtain the result advertised in Section \ref{sec:sumsad}, we take the semiclassical limit recalling that $\sigma_L\sim c$, and obtain

 \begin{equation}
     \mathcal{H}= \; \frac{\omega(\sigma_L)\;\;}{(\zeta_3\;\chi_4)^{2\sigma_L}}\;\,{}_2F_1(2\sigma_L,\, 2\sigma_L;\, 4\sigma_L;\, 1-\frac{\chi_3 \zeta_4}{ \chi_4\zeta_3})\;,
 \end{equation}\\[-4ex]

as claimed.

\section{Discontinuity of the Hypergeometric Function}
\label{sec:hypdisc}

In this section, we briefly review the discontinuity of the hypergeometric function, and derive an expression for the function on the second sheet as we cross the branch cut from above (c.f. Figure \ref{fig:contours}). Starting from the integral representation,

 \begin{equation}
 	\label{eq:intrep}
 	{}_2F_1 (a,b;c;x) = \frac{\Gamma(c)}{\Gamma(b)\Gamma(c-b)} \;\int_0^1 t^{b-1}(1-t)^{c-b-1}(1-tx)^{-a}\,dt \;\;,
 \end{equation}\\[-4ex]

where $\Real(c)>\Real(b)>0$, we note that the integral has a pole at $t=1/x$. The pole lies on the integration contour precisely when $x$ lies on the branch cut of the hypergeometric function $x\in[1,\infty)$. Let us therefore split the integral at the critical value of $t=1/x$:

 \begin{equation}
 \begin{split}
 	{}_2F_1 (a,b;c;x)\; =\; \frac{\Gamma(c)}{\Gamma(b)\,\Gamma(c-b)}\;\; \Bigg{\{} \;\;&\;\textcolor{gray}{\underbrace{\textcolor{black}{\int_0^{1/x} t^{b-1}(1-t)^{c-b-1}(1-tx)^{-a}\,dt}}_{\text{I}}}\\ 
  \;+\,&\; \textcolor{gray}{\underbrace{\textcolor{black}{\int_{1/x}^{1} t^{b-1}(1-t)^{c-b-1}(1-tx)^{-a}\,dt }}_{\text{II}}}\;\;\,\Bigg{\}}
\end{split}
 \end{equation}\\[-4ex]

As $x$ crosses the branchcut of the hypergeometric function from above, the first integral  $\text{I}$ is unchanged, whereas the second integral $\text{II}$ picks up a complex phase:

 \begin{equation}
 	\text{II}_+ \;\longrightarrow\; e^{2 \pi i \,a}\; \text{II}_-
 \end{equation}\\[-4ex]

 Here the subscripts indicate whether the argument $x$ lies above or below the branch cut. It is now possible to repackage the two terms separately into new hypergeometric functions of shifted weights, by suitably substituting the integration variable. For integral $\text{I}$ this is achieved by the substitution $t=s/x$ and introducing new labels $\alpha=1+b-c$, $\beta=b$ and $\gamma=b-a+1$:

 \begin{equation}
 \label{eq:one}
 \begin{split}
 	\text{I}\;
 			&=\;x^{-b} \;\int_0^1 s^{b-1} \left(1-\frac{s}{x}\right)^{c-b-1} (1-s)^{-a}\;ds\\[2ex]
 			&=\;x^{-b}\,\frac{\Gamma(\beta)\,\Gamma(\gamma-\beta)}{\Gamma(\gamma)}\;{}_2F_1(\alpha,\,\beta;\,\gamma;\,x^{\uminus 1})
 \end{split}
 \end{equation}\\[-4ex]
 
 Because in the limit of interest we want to set $x=z\!\rightarrow\!0$, for which the argument of the hypergeometric function above becomes singular, we rewrite the expression using identity (15.8.1) in \cite{822801}. Restoring the original weights $a$, $b$ and $c$, we obtain the final expression\\[-2ex]

 \begin{equation}
 	\text{I} \;=\;\frac{\Gamma(b)\,\Gamma(1\!-\!a)}{\Gamma(b-a+1)}\;\;x^{-b}\;(1-x^{-1})^{c-b-1}\;{}_2F_1(\,1+b\!-\!c,\,1\!-\!a;\,b\!-\!a+1;\,\frac{1}{1-x}\;)
 \end{equation}\\[-4ex]\;\;.

Similarly, we can repackage integral $\text{II}$ by making the substitution $t=(1-\frac{1}{x})s + \frac{1}{x}$, obtaining\\[-5ex]

\begin{equation}
 \begin{split}
 	\text{II} &= (-1)^{c-b}(1-x)^{c-a-b}\,x^{1-c}\,\int_0^1 ((x-1)s+1)^{b-1}\,(1-s)^{c-b-1}\,s^{-a}\;ds\\[0.5ex]
 	&=(-1)^{c-b}(1-x)^{c-a-b}\,x^{1-c}\;\frac{\Gamma(1\!-\!a)\,\Gamma(c\!-\!b)}{\Gamma(c\!-\!b\!-\!a+1)}\;{}_2F_1(1\!-\!b,1\!-\!a;c\!-\!b\!-\!a+1;1\!-\!x)\;\;.
 \end{split}
 \end{equation}\\[-3ex]

The full continued hypergeometric factor then simply reads\\[-2ex]

 \begin{equation}
 \label{eq:12}
 	{}_2F_1^{\text{\;II}}(a,b;c;x) = \;\frac{\Gamma(c)}{\Gamma(b)\,\Gamma(c-b)} \left\{ \,I \,+\; e^{2\pi i\, a}\;\text{II}\,\right\}\;.
 \end{equation}\bigskip

\section{Poisson Summation Formula}
\label{sec:poisson}
  
  Here we give a brief review of the Poisson Summation Formula and its application to the Dirac delta function. For this purpose, we define our Fourier transform using the following convention:

  \begin{equation}
  \hat{f}\;:=\;\int_{\mathbb{R}}\;e^{-2\pi ikx} f(x) \;dx
  \end{equation}\\[-2ex]

  Then, the following theorem holds.\\[-2ex]

  \begin{thm*}(Poisson Summation Formula)\\
  	Let $f:\mathbb{R}\rightarrow \mathbb{C}$ be a Schwartz function, and $\hat{f}$ its Fourier transform.
  	Then\\
  	 \[\sum_{n\in\mathbb{Z}}\;f(n)\;=\; \sum_{k\in\mathbb{Z}} \;\hat{f}(k)\]   
  \end{thm*}\medskip

  In particular, we can now apply this theorem to the special case of $f(x)=e^{2ni\xi}$. First, we determine the Fourier transform:
  \begin{equation}
  \begin{split}
  	\hat{f}(k)\;&=\;\int_{\mathbb{R}}e^{-2\pi ikx}\,e^{2xi\xi}\;dx\\[1.5ex]
  	&=\;\delta\Big(\frac{\xi}{\pi}-k\Big)
  \end{split}
  \end{equation}\\[-2ex]

  Then, the Poisson Summation Formula yields the following relation:

  \begin{equation}
  	\sum_{n\in\mathbb{Z}}\;e^{2ni\xi}\;=\; \sum_{k\in\mathbb{Z}} \;\delta\Big{(}\frac{\xi}{\pi}-k\Big{)}
  \end{equation}\\[-2ex]

 Setting  $\;\xi=-\tilde{\alpha}\pi/b^2\;$  recovers the claim in equation (\ref{eq:deltaid}) in Section \ref{sec:sumsad}.

\bibliographystyle{utphys}
\bibliography{LiouvilleChaos}

\end{spacing}
\end{document}